\documentstyle[12pt]{article}
\input epsf
\begin{document}
\title{
\hfill{\small PRA-HEP 97/06}
\\
\hfill{}
\\
\hfill{}
\\
\vspace*{0.5cm} \sc
On the role of power expansions in quantum field theory
\vspace*{0.3cm}}
\author{{\sc Jan Fischer}
%${}^1$\thanks{On}
\vspace*{0.3cm}}
\date{
\it Institute of Physics, Academy of Sci. of the 
Czech Republic, \\ Na Slovance 2, CZ-18040 Praha 8, Czech Republic}
\maketitle
\begin{center}
{\large\sc abstract} 
\end{center}
Methods of summation of power series relevant 
to applications in quantum theory are reviewed, 
with particular attention to expansions in powers
of the coupling constant and in inverse powers of an 
energy variable. Alternatives to the Borel summation 
method are considered and their relevance to different 
physical situations is discussed. Emphasis is placed on 
quantum chromodynamics. Applications of the renormalon 
language to perturbation expansions 
(resummation of bubble chains)  
in various QCD processes are reported and the importance of 
observing the full renormalization-group invariance in
predicting observables is emphasized. 
News in applications of the Borel-plane formalism to 
phenomenology are conveyed. 
The properties of the operator-product expansion along 
different rays in the complex plane are examined and the 
problem is studied how the remainder after subtraction 
of the first $n$ terms depends on the distance from 
euclidean region. Estimates of the remainder are 
obtained and their strong dependence on the nature of the
discontinuity along the cut is shown. Relevance of this 
subject to calculations of various QCD effects 
is discussed.  

\vspace*{0.2cm}

 
\vspace*{0.6cm}

\noindent April 1997\hfill

\thispagestyle{empty}
\baselineskip 20pt
\newpage

\setcounter{page}{1}

\tableofcontents

%%%%%%%%%%%%%%%%%%%%%%%%%%%%%%%%%%%%%%%%%%%%%%%%%%%%%%%%%%%%%%%%%%%%%%%%

%\noindent
\section{Introduction}
%{\bf 1.} {\em Introduction}.-
%%%%%%%%
\subsection{Perturbation series} 
%%%%%%%%%%%%%%%
A typical situation in physics is want of exact solutions. 
To find a good approximation of a problem, we have mostly first 
to simplify it by neglecting a number of effects, thereby 
facilitating its solution but, simultaneously, endangering 
its physical relevance. It is a general experience that 
exact equations can mostly be solved only approximately.

Perturbation theory (PT) is based on the idea 
of expressing the solution $f(g)$ to a problem (equation)  
in the form of the power series 
\begin{equation}
\sum_{n=0}^{\infty}f_{n} g^{n} ,
\label{Pt1}
\end{equation}
where $g$, the perturbative parameter, is considered to be a
small numerical quantity describing, possibly, a physically 
important effect. 

The first question to ask  
is how $f(g)$ is related to the series (\ref{Pt1}). 
To approach it, we have to see (i) what is the 
precise meaning of (\ref{Pt1}) and, (ii), in what sense the
series (\ref{Pt1}) is assigned to $f(g)$, the searched 
solution to the problem. For this to discuss, we have to 
consider complex values of $g$. %the coupling parameter $g$.  

Simplest is the situation when $f(g)$ is holomorphic inside a
circle of radius $\rho > 0$ centred at the origin; then
(\ref{Pt1}) is the Taylor expansion of $f(g)$ and determines
$f(g)$ uniquely, $f(g)$ being {\it equal} to (\ref{Pt1}) for all 
$|g| < \rho$. The $f_n$ are given  by the derivatives of $f(g)$ 
at the origin.

If, however, $\rho = 0$, the equality 
\begin{equation}
f(g)=\sum_{n=0}^{\infty}f_{n} g^{n} \,\, , 
\label{equal}
\end{equation}
being valid only for $g=0$, %loses its relevance 
is not relevant for applications.
On the other hand, if we assume, {\it instead} of %interpret 
(\ref{equal}), an {\it asymptotic} relation between $f(g)$ 
and (\ref{Pt1}),  
\begin{equation}
f(g) \sim \sum_{n=0}^{\infty}f_{n} g^{n} 
\label{asympt}
\end{equation}
for $g \rightarrow 0$, there is always an $f(g)$ satisfying 
(\ref{asympt}) for any set $\{f_n\}$ \cite{Hardy}. Further, 
in contrast to (\ref{equal}), the relation (\ref{asympt}) 
does not determine the function $f(g)$ uniquely, even when all 
the $f_n$ are explicitly given and the set of rays approaching
the point $g=0$ is specified. Even then a whole class of 
functions $f(g)$ is defined by (\ref{asympt}). The situation 
changes if additional restrictions are 
imposed on the remainder $f(g) - \sum_{n=0}^{N} f_{n} g^{n}$;
then there may be only one function satisfying the requirements,
and we can call it the sum of the series. 

Such additional restrictions as well as the assumptions 
(\ref{equal}), (\ref{asympt}) themselves should be 
motivated physically. Equation (\ref{equal}) is related to 
the question of convergence:
\begin{itemize}
\item /1/ (Convergence): Is the series convergent, 
in the sense that the limit of
\begin{equation}
f_{(N)}(g)=\sum_{n=0}^{N}f_{n} g^{n}
\label{Nsum}
\end{equation}
for $N\rightarrow \infty$, with $g$ fixed, exists? For what
values of $g$ is this the case?

The asymptoticity property (\ref{asympt}), on the 
other hand, is a statement on smoothness:
\item /2/ (Smoothness): Is the expression
\begin{equation}
|f(g) - \sum_{n=0}^{N} f_{n} g^{n}|
\label{rem}
\end{equation}
bounded, in what domain of $(N, g)$ (with $g$ 
complex and $N$ a nonnegative integer) and what is the 
bounding function?

Finally, there is the question of uniqueness:
\item /3/ (Uniqueness): Can the sequence $\{f_{n}\}$, 
$n=1,2$,...,  uniquely determine the expanded function 
$f(g)$ and under what conditions?
\end{itemize}  

Perturbation theory yields, at least in principle, explicitly 
all the coefficients $f_{n}$ of the expansion (\ref{Pt1}). 
The knowledge of the large-order behaviour of the $f_{n}, 
n=1,2$,..., tells us whether the series (\ref{Pt1}) is 
convergent or not, but what we really need is to answer the 
question /3/, i.e., that of a unique determination of $f(g)$.
If $f(g)$ is holomorphic at the origin, (\ref{Pt1}) is  
convergent and $f(g)$ is uniquely determined from the 
coefficients $f_{n}$ by means of (\ref{equal}). 
There was a universal belief till the early 1950's that a sound 
physical theory should not use a divergent perturbation 
expansion or a function singular at the  origin.\footnote{
This opinion survives among many physicists  till  the  present 
time. There was a similar mistrust of the use of divergent 
series also among mathematicians, but much earlier. J. 
Basdevant \cite{JB} gives  two interesting passages, by J. 
d'Alembert and by N.H. Abel. Other  examples illustrating the 
evolution of opinions on divergent series are in ref. 
\cite{Hardy}.} %with vanishing convergence radius.  

In 1952, F.J. Dyson \cite{Dyson} argued that in QED there is 
a singularity at the origin of the complex coupling-constant 
plane. (The conjecture that such a singularity
makes the series divergent was 
questioned later, see \cite{PMS} and references therein.) 

A wide class of divergent series can be given precise
meaning by means of Borel summation, provided
that certain additional conditions are imposed on $f(g)$. This
method has found fruitful applications in quantum field 
theory, where its use goes back to the 1950's. In 1951, for 
example, J. Schwinger \cite{Schwi} obtained a compact 
expression for the term in %addition $L'$ to 
the Lagrangian corresponding to
the QED vacuum polarization by an external constant magnetic
field $H$. Then  V. Ogievetsky \cite{Ogi} used the (divergent) 
perturbative expansion of this term %$L'$ 
to show that its Borel sum 
coincides with Schwinger's result. Indeed, the coefficients of 
its perturbative expansion %coefficients of $L'$ 
in powers of $e^2$ behave as 
$(2n-3)! (-H^{2}/(\pi^{2}m^{4}))^n$ at large $n$, where $e$ and 
$m$ is the charge and the mass of the electron respectively. 
Ogievetsky found the Borel sum in the form
\begin{equation}
\int_{0}^{\infty} {\rm e}^{-m^2 t} a(t,e) {\rm d}t
\end{equation}
with
\begin{equation}
a(t,e)=-\frac{1}{8 \pi^2 t^3}[etH\cot(etH)-1-(etH)^2 /3],
\end{equation}
which coincides with the compact expression obtained by
Schwinger \cite{Schwi}. This important result shows that a
divergent perturbative expansion does not signal an
inconsistency in a theory; it also shows that there are
special - but realistic - cases of Borel summability in QED, 
although general considerations indicate Borel non-summability
(see \cite{Azim,'t Hooft,'t Khuri}, and Table 1 and a 
discussion in section 2 of the present paper).  

Gradually, the Borel summation techniques became widely adopted 
in quantum theory. It even became common %usual a fashion  
to {\it judge} theories according  to whether their 
perturbation expansions are Borel summable or not; in this 
sense, the formal criterion of Borel summability became 
successor to the criterion of convergence. 
Being rather technical assumptions, however, neither 
convergence nor Borel summability can serve as criteria of  
consistency of a theory. 
One might find it more natural to set another 
criterion of consistency: 
that the transition from the perturbed to the unperturbed
state should be continuous  and  smooth in  the  coupling 
parameter. But even this 
is not necessary: Nature is not obliged to be smooth in our
parameters.
It is a special step to assume that the transition from 
the world with interaction, $g>0$, to that without it 
($g=0$) is smooth. This is why the relation
(\ref{asympt}) has nontrivial %and far-reaching 
physical implications. 

Disregarding special field theories with interactions being 
interpreted  
as discontinuous perturbations in the sense that 
the theory does not return to the unperturbed theory as 
the interaction coupling vanishes,  
the fact is that asymptotic expansions in powers 
of $g$ have been universally adopted in quantum theory; 
perturbation theory does imply the assumption of a smooth 
transition from the state with interaction 
to that without it. What we then really need is to answer 
the question /3/ of uniqueness, which is independent both 
of convergence and of Borel summability. If the answer 
to /3/ is yes, i.e., if the $f_{n}$ determine $f(g)$  
uniquely, the 
theory could be considered as selfsufficient, no matter
whether the conditions of {\it Borel} summability are 
satisfied or not.  If the answer is no, the theory has to 
be supplemented with some additional input. 

In the majority of situations in physics, $f(g)$ has a 
singularity at the origin; note that smoothness at a point 
does not exclude a singularity at this point. 
Then the perturbation expansion is not a Taylor 
series and, if considered as an asymptotic expansion,  
it needs a supplement %to reach uniqueness. 
for uniqueness to be reached. Needless to say, this 
supplement should be of {\it physical} nature, while no  
formal argument (e.g., that the method yields a finite 
value of a divergent series) can fill the gap where physics 
is missing.

%%%%%%%%%%%%%%%%%%%%%%%%%%%%
\subsection{Large-order behaviour}

The problem of the high-order behaviour of the 
perturbation-expansion coefficients in field theory 
calculations has aroused new interest in the last  
decade. 
Particular attention has been paid to power corrections 
to QCD predictions for hard scattering processes. One can 
point out two reasons for this growing interest:
\begin{itemize}
\item The theoretical problem of how physical observables can be
  reconstructed from their (often divergent) power expansions.
  
\item The pragmatic need to assess the usefulness of performing the
  extensive evaluations of multi-loop Feynman diagrams in QCD. Much
  effort has been devoted to the computation of higher-order QCD
  perturbative corrections; in some cases third-order approximations
  are known, and we now seem to be at the border of what can be
  carried out analytically or numerically in high-order perturbative
  calculations. So, why next order? If the series is divergent, 
  the next order may represent no improvement with respect to the 
  lower-order result. On the contrary, at a certain
  order it will lead to deterioration.

\end{itemize}

The question of the possible divergence of %the corresponding 
{\it asymptotic} perturbation expansions, which goes back to 
the original argument of Dyson \cite{Dyson}, still defies 
definite and clear answer. Notice 
in this connection that a series that is asymptotic to a 
function singular at the origin need not be divergent. 
Indeed, for any series (\ref{Pt1}) convergent within a 
circle $|g|<\rho$ there are functions $f(g)$ which are singular 
at $g=0$ and have (\ref{Pt1}) as asymptotic series. The traditional wisdom 
that the divergence of 
perturbation expansions is directly related to the
discontinuities of Green's functions is not generally true.  
If the series is asymptotic, the knowledge of the large-order
behaviour of perturbation-theory coefficients is insufficient
for the determination of the function expanded and one needs
much more detailed information in the vicinity of the point of
expansion in order to establish the relation between the
expansion and the function expanded.  

In fact most of the information on the analytic properties 
of relevant Green's functions near $g=0$ \cite {'t Hooft, 
Mueller} may actually be invisible in perturbation 
expansions themselves. Moreover,  
the summability of a perturbation expansion is not the same as 
the possibility of recovering from such series the corresponding 
full physical quantity. %, has gained wider acceptance. 
As the summability of QCD perturbation expansions is 
determined by the large-order behaviour of their 
coefficients, we primarily need to know which kind of 
information is necessary to explicitly
obtain the sum of the series.   
 
If a theory provides, in addition to the values of the expansion 
coefficients ${f_{n}}$, enough supplementary information,   
a power series can be summed to a finite number, which is 
not arbitrary but reflects %the use of 
the additional piece of information used. One has to give it
such a form %The task of the summation method is to give 
that it could be used to reduce (or even remove) the 
ambiguity connected with the asymptoticity of the expansion. 
The task of a summation method then is not to 
assign arbitrarily an ad hoc value (whose only merit might 
just be that it is finite) to your infinite 
or ambiguous series, nor to fill 
the gap in physics (which your 
theory possibly possesses)  with some mathematical 
cement,  but to offer you a service: to translate 
your {\it physical} idea (or that of the theory you work 
with) into a mathematical condition to reduce the 
ambiguity of the formal power series. 

As a typical example of this procedure the role of  
general principles in the $S$-matrix theory can be 
mentioned. 
The $S$-matrix techniques were built from mathematically 
formulated, but physical principles: analyticity as a 
mathematical formulation of microscopic causality, 
unitarity as probability conservation, crossing 
symmetry as a general property of Feynman diagrams, and the 
position and character of the cut in the complex energy 
plane as a description of the properties of the particle 
spectrum. A simultaneous use of the analyticity 
properties and of the concepts of field theory was 
carried out in the QCD Sum Rules program \cite{SVZ}, 
which met with tremendous phenomenological success. Note 
that the foundation of this approach rests 
on the non-trivial fact \cite{OehZim, Oeh} that 
the principles of $S$-matrix theory are 
rigorously true in QCD: it follows \cite{Oeh} 
that for gauge theories, if confinement is imposed, the 
analytic properties of physical amplitudes are the same 
as those obtained from an effective theory involving 
only the composite (physical) fields. 

It should be considered as very fortunate that, 
simultaneously, analyticity 
plays a crucial role also as a {\it mathematical} condition 
reducing the ambiguity of asymptotic series. In Section 2
of the present paper, we discuss in detail the interplay 
between large-order behaviour of a series (as listed in 
Table 1) and the analyticity properties of the function 
expanded; it turns out that a balance 
between these two concepts is needed  
for a unique determination of $f(z)$ from 
(\ref{asympt}), in the sense that if more 
analyticity of $f(z)$ is available, one can afford a more 
violent behaviour of the $a_{n}$, and vice versa. 

In Section 3 we focus on some practical aspects of the 
operator-product expansion, in particular on the problem
of how the remainder after subtraction of the first $n$ 
terms from the function expanded depends on the distance 
from euclidean region, provided that an estimate on the 
remainder in euclidean region is known. Sections 4 and 5 
are devoted to singularities in the Borel plane and 
applications of the Borel plane formalism to phenomenology. 
We discuss the progress in understanding the role of the 
singularities in the Borel plane for nonperturbative 
effects, in studying hadrons containing heavy quarks, then 
the different methods of calculation of the coupling 
constant $\alpha_{s}$ from existing experimental data, 
the role of operator product expansion in practical
calculations  and the progress in the resummation of 
bubble chains. Section 6 is devoted to problems related 
to renormalization ambiguities and their impact on
phenomenology. Section 7 contains concluding remarks.  
Some parts of the present paper represent an extended and amended
version of ref. \cite{zako}.   
       
Below we give a survey of the large-order behaviour of
perturbative series in different theories. 
Bender and Wu \cite{Bender-Wu} examined  
perturbation theory for one-dimensional quantum mechanics 
with a polynomial potential 
and they obtained the large-order 
behaviour of the perturbation coefficients. 
Lipatov \cite{Lipatov} obtained the same results for 
massless renormalizable scalar field theories.

Br\'ezin, Le Guillou and Zinn-Justin \cite{Brezin}, applying
independently the same method to anharmonic oscillations in 
quantum mechanics, were able to rederive and generalize the 
results obtained by Bender and Wu.

In QED, after the pioneering work of Dyson \cite{Dyson}, a 
number of papers appeared discussing the analyticity properties 
of the Green functions (some of them are cited in refs. 
\cite{PMS} and \cite{LeGuillou}). The growth like $n!$ 
(where $n$ is the order of approximation) has two sources:

1. The number of diagrams grows like $n!$, each diagram giving 
a contribution of the order of 1.

2. There are types of diagrams for which the amplitude itself 
grows like $n!$ \cite{GrN}.

A survey of the large-order behaviour of expansion 
coefficients in some typical theories and models is given in 
the Table 1. As subtle cancellations among higher-order 
graphs may occur, the expressions in the third column of 
Table 1 may sometimes give 
an upper bound rather than
the actual high-order behaviour of the coefficients. 

Table 1 is intended for first information and should not 
be used for systematic analyses because some important 
conditions or restrictions could not be mentioned. 
A brief explanation of its use is given below.  
Notation is not consistent, being mostly taken from the 
papers quoted; the reader is referred to them for explanation. 
References in the table as well as in the text are often made 
not to the original papers but rather to more recent papers or 
reviews, in which the subject is explained in modern context.
As a result, a number of valuable papers are not quoted. I 
apologize for this. For details %the reader is referred 
I refer to the papers cited and references therein, and to 
the anthology by Le Guillou and Zinn-Justin \cite{LeGuillou}, 
which contains a list of references up to \,1990.

\begin{table}
\begin{tabular}{lll}
{\bf Theory}  & {\bf Notation and references} & {\bf High-order 
behaviour} \\[10pt]
\phantom{} 1 Disp. relation & $ {\rm Im}f(z) \sim z^{-b} 
{\rm e}^{-a/z}$ & $f_n \sim (-a)^n \Gamma (n+b)$  \\
\phantom{} 2 Anharmonic  & $E_{m}(g) = \sum_{n=0}^{\infty} E_{m,n} 
g^n$ & $E_{m,n} \sim (-3/4)^n n!$ \\ \phantom{2 } oscillator 
& \cite{Bender-Wu} \\
\phantom{} 3 Anharmonic  & $I(g) \sim \int_{-\infty}^{\infty} 
{\rm e}^{(-x^2 /2 - gx^4 /4)} {\rm d}x$ & 
$I_n \sim (-1)^n 4^{-n} (n-1)! $ \\ \phantom{3 } oscillator, 
\,\, $\phi^4$ & \cite{LeGuillou} \\
\phantom{} 4  $\phi^4$ in 2 and & \cite{EcMaSe} & Analyticity for
complex $g$, \\ \phantom{4 } 3 dimensions & \cite{MagSen} & Borel
summability proved \\

\phantom{} 5 Instantons, & $m \geq 6$, \cite{Lipatov} %BogomFat1} 
& $C_{n}(m) 
\sim (-\tilde{g})^n {\rm e}^{n(1-m/2)} n^{(m+D)/2}$  \\
\phantom{5 } $D=2m/(m-2)$
& $m=4$,  \cite{Lipatov} & $C_{n}(4) 
\sim 2.75n^{4}(n/(16 \pi^2 e))^n $ \\
\phantom{} 6 Field theories & $Z_{n}=\sum_{k=0}^{\infty} 
Z_{n}^{(k)}(-e^{2}/(4\pi))^k$ & $Z_{n}^{(k)} \sim k! 
k^{b_{n}} A^{-k} c_{n}[1+O(1/k)]$ \\ 
\phantom{6} without fermions &
\cite{Brezin},\cite{Itzykson} - \cite{Hurst}\\
\phantom{} 7 Field theories & \cite{Parisi1} & $A_{n} \sim \Gamma 
(n (d-2)/d) R^{n} n^{-\alpha} \cos(2\pi n/d) A$ \\
\phantom{7 } with fermions & \cite{BogomFat2} & $f_n 
\sim (-1)^n S^{n/2} \Gamma(n/2)D$ \\
\phantom{} 8 Yukawa theories & \cite{Parisi1} & $Z_{n} \sim n^{-\alpha} 
A^{-n} {\rm cos}(2 \pi n/d) \Gamma (n (d-2)/d)$ \\
\phantom{Yuk} $d=2$ & \cite{Fry} & $d_{n} \sim A^{-n} 
(\ln\, n)^n $ \\
\phantom{} 9 QED & \cite{BogomFat2,Balian} 
& $f_{n} \sim (-1)^{n} \Gamma((n+\nu)/2) a^{n+\nu}$ \\
& \cite{BenZakh} (1st IR renormalon) 
& $d_{n} \sim {\rm const}
(-\beta_{0}/2)^{n} \Gamma(n+1-2\beta_{1}/\beta_{0}^{2})$\\
& \cite{BenSmi} & $r_{n} \sim {\rm const} \, \beta_{0}^{n} n! \,
n^{2+\beta_{1}/\beta^{2}_{0}+\gamma_{2}}$ \\ %11/8}$ \\
10 QCD in 2 dim. & \cite{Zhit} & $(g^{2}N_{c}/2)^{2k}(-1)^{k-1}(2k)!$
\\ 
11 QCD & \cite{Mueller,Vain,Beneke1,Zakh} & $f_{n} \sim a^{n} n^{\gamma} n!$ 
\\ 
& see \cite{BenZakh,BenSmi} for discussion & $d_{n} \sim {\rm const}
(-\beta_{0}/2)^{n} \Gamma(n+1-2\beta_{1}/\beta_{0}^{2})$\\
12 Bosonic strings  & $h$ is \# of handles,\,\, \cite{Periwal} & 
$\sim h!$ 
\end{tabular}
\caption[]{{\bf High-order behaviour of perturbation expansion 
coefficients} \\
(see also explanations in the text)}
\end{table}

A few words on the explanation of the symbols used in the
table. In the item 1, $f(z)$ is a real function analytic in 
the complex $z$ plane cut along $[-\infty, 0]$ and equal 
to $\int_{-\infty}^{0}{\rm d}z'{\rm Im}f(z')/(z'-z)$. In 
item 2, the $E_{m}(g)$ are, for $g>0$, the eigenvalues of the 
Hamiltonian $H=-\frac{1}{2}({\rm d/d}q)^2 + \frac{1}{2}q^2 
+ \frac{1}{4}gq^4$.

Eckmann, Magnen and S\'en\'eor \cite{EcMaSe} consider the
two-dimensional Euclidean boson field theories and give bounds on
truncated Schwinger functions. They also prove that the domain of
analyticity and of the bounds obtained can be extended to an angle
$> \pi$ bisected by the positive real semiaxis, and prove the Borel
summability of the power series of the Schwinger functions. In the
case of the Euclidean $\phi^4$ theory in 3 dimensions, Magnen and
S\'en\'eor \cite{MagSen} prove the stability of the free energy for
complex values of the coupling constant and obtain, in analogy 
with the two-dimensional case, Borel summability of the
perturbation series.  

In the instanton row (item 5), the $C_{n}(m)$ are the coefficients 
of the expansion of the Gell-Mann--Low function, with $n$ being 
the perturbative order and $m$ being related to the dimension $D$ 
by $D=2m/(m-2)$. Further, both $C_{n}(m)$ and  $C_{n}(4)$ 
contain an $n$-independent (but $m$-dependent) factor. 

The item 6 is related to scalar electrodynamics in 
four dimensions, with the action
\begin{equation}
S(\phi, A_{\mu})=\int{\rm d}^4 x\left[\frac{1}{4}F_{\mu 
\nu}^{2}+\frac{1}{2}|D_{\mu}\phi|^2 +\frac{1}{2}m^2 |\phi|^2 + 
\lambda |\phi|^4 /4! \right] .
\end{equation}
The $Z_{n}$ are one-particle irreducible Euclidean Green's 
functions, with $c_{n}, b_{n}$ and $A$ being  constants and 
$A$ being equal to 12 approximately. 

For field theories with fermions (item 7), $A(g) = 
\sum_{n}A_{n}g^{2n}$ is the Green function of the scalar field 
$\sigma$ with the coupling term $g\sigma \bar{\Psi}\Psi$ for a 
Yukawa interaction  in  $d>2$  dimensions  \cite{Parisi1};  $R, 
\alpha$ and $A$ are computable constants. In ref.  
\cite{BogomFat2}, the $f_n$ denote the main contribution to the 
QED perturbation theory asymptotics, where $S$  is 
equal to $3^{-3/2} 4\pi^3$.

The $d_n$ in the item 9 describe (see \cite{BenZakh}) the 
asymptotic behaviour of perturbation series due to the first 
infrared renormalon in QED. (In QCD, item 11, the result agrees 
with that obtained by relating infrared renormalons to 
nonperturbative corrections.) The $d_{n}$ are the coefficients 
in the perturbative expansion of the function $Q^2 
{\rm d}/({\rm d}Q^2) \,\, P(Q^2)$ in powers of the coupling 
parameter $\alpha(Q)$, 
\begin{equation}
\sum_{n=-1}d_{n} \alpha (Q)^{n+1} .
\label{dn}
\end{equation}
In QED, $\alpha P$ is equal to $\Pi$, the 
photon vacuum polarization, and $d_{-1} = - \beta_{0}$, 
$\beta_{0}$ is the first coefficient in the expansion of 
the beta function, which is defined as $\mu^2 {\rm d}/{\rm d} 
\mu ^2 \,\, \alpha(\mu) = \beta(\mu)= \sum_{n=0} \beta_{n} \alpha 
(\mu )^{n+2}$. In QCD, $P=\Pi$ is the electromagnetic
current-current correlation function. The large-order behaviour 
in QCD is %\cite{Mueller, Zakh} 
$d_{n} \sim {\rm const}(-\beta_{0}/2)^{n} 
\Gamma(n+1-2\beta_{1}/\beta_{0}^{2})(1+{\rm O}(n))$. 
Beneke and Zakharov \cite{BenZakh} show that in QED 
(with massless fermions), the coefficients $d_n$ have the 
same IR behaviour with $\beta_{n}$ taking their QED values. 
The UV renormalons are also present and 
produce the fixed-sign divergence of the perturbative 
series. They have been analyzed by Beneke and Smirnov
\cite{BenSmi}, who study classes of diagrams with an arbitrary
number of chains, and derive explicit formulae for leading and
subleading divergence. To organize the diagrams in classes, the
expansion parameter $1/N_{f}$ is used, where $N_f$ is the
number of fermion species; as a consequence, diagrams
suppressed in the $1/N_f$ expansion are not suppressed for
large $n$ and, consequently, no finite order in the $1/N_f$
expansion provides the correct behaviour in $n$ in the full
theory. Table 1 shows the large-order behaviour of the 
vacuum polarization, $r_{n}$ being the coefficient of 
i$\alpha^{n+1}$ in the perturbative expansion and 
$\gamma_{2}=99/(8 N_{f}^{2})$ . The authors 
discuss extension of the formalism to non-abelian 
gauge theories and expect a similar result.  

For QCD in 2 dimensions (with $N_{c}$ tending to infinity), 
Zhitnitsky \cite{Zhit} argues that the factorial growth 
$(g^{2}N_{c}/2)^{2k}(2k)!(-1)^{k-1}$ of the large-order 
perturbative coefficients is related to instantons, 
thus being of non-perturbative nature. Important in this 
theory is dimensionality of the coupling constant.     

Gross and Periwal \cite{Periwal} (see item 12) proved that the 
$h$-loop bosonic partition function (cut off in the 
infrared region) is bounded below %by $k'k^{h}h!$ 
and the perturbative 
expansion for the bosonic string diverges as 
$\sum_{h=0}^{\infty} g^{2h}\,h!$. The series is not
Borel summable, all its terms being positive.

A look at the third column of Table 1 shows that most of the theories
listed are characterized by an $n!$ large-order behaviour. 
{\it This does not mean that all of them can be cured by the 
same resummation method}: large-order behaviour is just one 
of aspects which determine the summation procedure. To each 
power series with coefficients listed in the 3rd column of Table 
1, there is a whole class of functions $f(z)$ having the same 
asymptotic expansion. To specify the asymptotic expansion, one 
has to establish the angle (ray(s)) along which $z$    
approaches  the origin; 
further, to pick  out one function $f(z)$ of this class, 
one has to add some additional information, %specifying 
according to the theory in question. 

As was already mentioned, these additional conditions are often 
formulated in terms of analyticity properties of the function 
expanded. 

In the next section, let us discuss the character of the additional
conditions.

%%%%%%%%%%%%%%%%%%%%%%%%%%%%%%%%%%%%%%%%%%%%%%%%%%
%%%%%%%%%%%%%%%%%%%%%%%%%%%%%%%%%%%%%%%%%%%%%%%%%%
\section{Asymptotic series. Summation methods}

\begin{flushright}
If you are in doubt, \\
examine the asymptotics
\end{flushright}

How to deal with divergent series and under which conditions a power
series can uniquely determine the expanded function are questions of 
fundamental importance in quantum theory. Power expansions are 
badly needed in physics, but additional input is required to 
ensure that they have precise meaning. These additional conditions 
should reflect some physical features of the system.

We shall change the notation in this section, introducing the symbol
$z$ for the coupling parameter denoted previously by $g$. 

A formal series (\ref{Pt1}) is called asymptotic to $f(z)$ on 
the set ${\cal S}$ if the sequence $R_{N}(z)$,
\begin{equation}
R_{N+1}(z) = f(z) - \sum_{0}^{N} f_{n}z^{n}
\label{R}
\end{equation}
satisfies the condition
\begin{equation}
R_{N+1}(z) = o(z^{N})
\label{Ro}
\end{equation}
for all non-negative integers $N$ and all $z \,\in 
\,{\cal S}$, where ${\cal S}$ is a subset of the complex 
plane having the origin as an accumulation point.

The analogous notion of a series for which (\ref{R}) and
(\ref{Ro}) are satisfied only for a %finite//
limited number of $N$, $N=0,1,2,...,N_{0}$, has also
interesting applications.

It is important that ${\cal S}$ is generally a subset 
of the neighbourhood of the origin. As an example to 
illustrate non-uniqueness, consider (\ref{asympt}) for $z$
approaching zero within the angle $|\arg z| < \beta$ with 
$\beta \leq \pi/2$, i.e., in the right-hand half of the complex 
plane. To a function $f(z)$ satisfying (\ref{asympt}), one can 
add a term of the form $A \exp (-1/z^{a})$ with $A$ real and 
$0<a<\pi/(2 \beta)$ without violating (\ref{Ro}) within 
${\cal S}$. 

A summation method states conditions under which an asymptotic
series can determine $f(z)$ uniquely, and yields the explicit 
formula for $f(z)$. 

%%%%%%%%%%%%%%%%%%%%%%%%

\subsection{Uniqueness by Borel summation}

The series 
\begin{equation}
\sum_{n=0}^{\infty}a_{n}z^{n}
\label{Pt2}
\end{equation}
is called Borel summable if 

1) its Borel transform $B(t)$,
\begin{equation}
B(t) = \sum_{n=0}^{\infty} a_n t^n /n! ,
\label{BT}
\end{equation}
converges inside some circle, $\mid t \mid < 
\delta \, , \,\, \delta > 0$;

2) $B(t)$ has the analytic continuation to an infinite strip
of non-vanishing width bisected by the %neighbourhood of the
positive real semi-axis ${\rm Re} \,\,t \geq 0$, and

3) the integral
\begin{equation}
g(z) = \frac{1}{z} \int_{0}^{\infty} {\rm e}^{-t/z} B(t) {\rm d} t  \, ,
\label{BS}
\end{equation}
called the Borel sum, converges for some $z \neq 0$.

One can see the motivation of this 
summation method on a simple example. 
Consider a generic quantity $D$, calculated 
in perturbation theory with the coupling $z$, 
\begin{equation} 
D(z) =  \sum_{n=0}^{\infty}d_n z^n \, . 
\label{1}
\end{equation}
This can be rewritten as
\begin{equation}
D(z) = \sum_{n=0}^{\infty}d_n z^n (1/n!) 
\int_{0}^{\infty}{\rm d}t {\rm e}^{-t}t^n \, .
\label{2}
\end{equation}
If the series (\ref{1}) has a non-vanishing convergence 
radius $\rho$, the integration in (\ref{2}) can be exchanged 
with the sum inside the circle. If we are outside the circle 
(or if the convergence radius is zero, $\rho =0$) we 
exchange the order of integration and summation to {\it 
define} the series by the same expression (provided that 
the integral converges). In either case, we obtain
\begin{equation}
D(z) =\int_{0}^{\infty}{\rm d}t{\rm e}^{-t} \sum_{0}^{\infty}
d_n \frac{(zt)^n}{n!}
=\int_{0}^{\infty}{\rm d}t{\rm e}^{-t}B(zt)  \,,
\label{3}
\end{equation}
where $B$ is the Borel transform of $D$. Taking $d_{n}= n!$ 
as an example (finite-order coefficients are irrelevant for 
the character of singularities) we obtain
\begin{equation}
D(z) =\int_{0}^{\infty}{\rm d}t{\rm e}^{-t}\frac{1}{1 - zt} \,.
\label{4}
\end{equation}
This integral does not exist for $z$ positive (where $D(z)$
has a cut), nor is the Borel sum of such a series defined. 
For other values of $z$, the 
integral is convergent and selects one of the functions 
having $\sum_{n=0}^{\infty}n!z^{n}$ as asymptotic expansion 
within the angle $0 < \arg z < 2 \pi$. The summation can be 
defined in many ways; there are infinitely many functions 
with this  asymptotic expansion. For instance, one can add 
any term of the form $A \exp(-1/z^{\alpha})$ for 
$0 < \alpha < 1/2$.

Comparing the expansions (\ref{Pt2}) and (\ref{BT}) we observe 
that the Borel transform has better convergence properties. 
To see this, let us compare the convergence radius  
$\rho_{1}$ and $\rho_{2}$ of (\ref{Pt2}) and (\ref{BT}) 
respectively. We have 

\begin{eqnarray}
1/\rho_{1} = \limsup_{n \rightarrow \infty} 
\sqrt[n]{|a_{n}|}  \\
1/\rho_{2} = \limsup_{n \rightarrow \infty} 
\sqrt[n]{|a_{n}|/n!} .
\label{rho}
\end{eqnarray}
The factor $\sqrt[n]{n!}$ between $\rho_{1}$ and $\rho_{2}$ 
indicates that if the convergence radius of the original 
series is nonvanishing, that of $B(t)$ will be infinite.
Also, some singularities that are condensed at the origin
will spread out %for $B(z)$ 
to the complex plane. In a sense, the distribution of 
singularities of the Borel transform 
is a blow-up of that of the original function; the 
convergence properties of (\ref{BT}) are  better than those 
of (\ref{Pt2}).

Nevanlinna \cite{Nevan} gave the following criterion of Borel
summability (this criterion is a refinement of the Watson
lemma \cite{Watson}, see \cite{Sokal} and also \cite{Fischer} 
for discussion):

Let $f(z)$ be analytic in ${\cal K}(\eta)$ defined 
by the inequality ${\rm Re}\frac{1}{z} > \frac{1}{\eta}$ 
(with $\eta$ positive), a disc of radius $\frac{1}{2}\eta$ 
bisected by the positive real semi-axis and tangent to the 
imaginary axis (see Fig. 1a), and let $f(z)$ have the 
asymptotic expansion 
\begin{equation}
f(z) \sim \sum_{n=0}^{\infty}a_{n}z^{n} . 
\label{as}
\end{equation} 
If the remainder $R_{N}(z)$ 
(\ref{R}), after subtracting $N-1$ terms from $f(z)$,
is bounded by the inequality
\begin{equation}
\vert R_{N}(z) | < A \sigma^{N} N!  |z|^{N} 
\label{bound}
\end{equation}
uniformly for all $z \, \in \, {\cal K}(\eta)$ and all $N$ 
above some value $N_{0}$, then $f(z)$ can be represented by 
the integral (\ref{BS})  
for any $z \, \in \, {\cal K}(\eta)$. 

Taking QCD as an example, we see that the conditions of 
the Nevanlinna criterion are not satisfied,
because the analyticity region is not a disc ${\cal K}(\eta)$ 
but a wedge of zero opening angle.\footnote{This 
non-perturbative result (see \cite{'t Hooft} 
and \cite{'t Khuri}) is obtained by combining asymptotic
freedom with analyticity and unitarity of two-point Green 
functions in the complex momentum squared plane; see 
also\cite{Azim}.}
 
Let us therefore consider some modifications of the 
Nevanlinna criterion.

%%%%%%%%%%%%%%%%%%%%%%%%%%%%%%%

\subsection{Analyticity vs. large-order behaviour: 
balance  for uniqueness}

We shall briefly mention two modifications of the Borel 
method that are relevant for quantum theory. There must be a 
balance between large-order behaviour and ${\cal A}$, the 
analyticity and boundedness domain of $f(z)$. %They also  
The rule is simple: if the expansion coefficients grow 
faster than $n!$, one has to require a larger %analyticity 
domain ${\cal A}$ to obtain a unique function from the 
expansion. If the large-order behaviour of the coefficients 
is tamer than $n!$, one can afford a smaller domain ${\cal
A}$ to reach uniqueness in determining $f(z)$. 

The Nevanlinna theorem can be generalized to the case that
the opening angle of the analyticity domain is less than
$\pi$, in which case the disc ${\cal K}(\eta)$ transforms 
into a drop-shaped domain with its tip at the origin. 

Indeed, let $F(z)$ be holomorphic in the "droplet" ${\cal L}
(\rho)$ placed inside the sector $|\arg z| \leq \frac{1}
{2}\pi \gamma$, $\gamma < 1$  
%the boundary of ${\cal L}(\rho)$ being defined by
%\begin{equation}
%|\sum_{n=0}^{\infty}z^{n}/\Gamma(\rho n+1)| = {\rm const} 
%\label{drop}
%\end{equation}
with $\rho \leq \gamma$. We demand, in analogy with 
(\ref{R}) and (\ref{bound}), the bounds in the form
\begin{equation}
|f(z)-\sum_{n=0}^{N-1} a_{n}z^{n}| \leq
A\sigma^{N}(N!)^{\rho}|z|^{N} . 
\label{boundr}
\end{equation}
The generalized Borel transform
will have the form
\begin{equation}
B_{\rho}(t) =\sum_{n=0}^{\infty}a_{n}t^{n}/\Gamma(\rho n+1) .
\label{BTrho}
\end{equation}
We see that the condition (\ref{boundr}) is {\it more
restrictive} than (\ref{bound}) but, in compensation, its
validity is required within a {\it smaller} angle. 
Correspondingly,
the coefficients of the series $B_{\rho}(t)$ defined by
(\ref{BTrho}) are {\it less suppressed} than those of $B(t)$
defined by (\ref{BT}) (notice that $B_{\rho}(t)$ equals
$B(t)$ for $\rho =1$).

In discussing this generalized form for different values of
$\gamma$ and comparing it with the original Nevanlinna 
theorem, we see that there is a balance between 
the size of the opening angle of ${\cal A}$ 
and the restrictiveness of the 
corresponding condition required for the series to 
determine the expanded function uniquely. This also 
means that Borel summability (i.e., $\rho = 1$, $\mu(n)
= n!$)) plays no privileged role among the variety of 
possible summation methods.

In many practical problems, 
the Borel method nevertheless seems to be preferable, 
because most of the 
large-order estimates suggest an $n!$ behaviour of the
perturbative coefficients (see Table 1). But this 
method simultaneously requires analyticity and the bound
(\ref{bound}) in the $z$ plane in an opening angle that 
is equal to $\pi$. This is not always satisfied; in QED and 
QCD, as was already mentioned,  
the opening angle is zero in the coupling-constant complex 
plane, the origin being an accumulation point of 
singularities, and ${\cal A}$ %the analyticity domain 
being a wedge bisected by the positive real semiaxis and 
bounded above and below by circles that are tangent to it. 

This generalization of the Nevanlinna theorem gives no answer 
to such a situation; a zero opening angle would make us 
choose $\gamma$ and $\rho$ equal to zero, but in this case 
the series (\ref{BTrho}) for $B_{\rho}(z)$ would coincide 
with the original one  
(\ref{Pt2}) and there would be no resummation. Conditions 
for a unique determination of $f(z)$  for ${\cal A}$  
of such a shape have been obtained by A. Moroz 
\cite{MorozCMP}. They are given by the following theorem. 

{\it Theorem} 1 \cite{MorozCMP}. Let $f(z)$ 

1) be regular in the wedge ${\cal W}$ \, with the boundary \,\, 
$|F(1/z)|=F(1/R)$,\,\, $F(z)=\sum_{n=0}^{\infty} z^{n}/\mu(n)$, with 
\begin{equation}
\mu(n)=\int_{o}^{\infty} \exp(-{\rm e}^{t})t^{n}{\rm d}t,
\label{mu}
\end{equation} 
and continuous up to the boundary. 

Let the remainder $R_{N}(z)$, see (\ref{R}), satisfy the bound 
\begin{equation}
|R_{N}(z)|\leq A\mu(N)|z|^{N} %\exp[-{\rm e}^{w(N)}+ N \ln w(N)]
\label{boundM} 
\end{equation}
uniformly in $z\,\in \,\bar{\cal W}$ and $N$. 

Then
\begin{equation}
M(t)=\sum_{n=0}^{\infty}f_{n}t^{n}/\mu(n)
\end{equation}
converges for $|t|<1$ and $f(z)$ is uniquely represented by the
absolutely convergent integral
\begin{equation}
f(x)=\int_{0}^{\infty}\exp(-{\rm e}^{t}) M(tx){\rm d}t
\end{equation}
for any $x \in (0,R)$. 

The interested reader is referred to the original work 
\cite{MorozCMP} for details and also to a discussion in  
\cite{Fischer}. We just mention that, according to this 
theorem, the horn-shaped (zero angle) analyticity 
domain requires, for $f(z)$ to be uniquely determined from 
the asymptotic series, a very tame large-order behaviour 
(given by (\ref{boundM})) 
of the expansion coefficients, namely $a_{N} \sim 
(\ln N)^{N}$ instead of $N!$ required in the ordinary Borel 
summation method. Note the double exponential function 
in the integrand of (\ref{mu}), which causes this slow
large-order behaviour.

\begin{table}
\begin{tabular}{llll}

{\bf $f(z)$ at $z=0$}  & {\bf Uniform bound} &  {\bf Transform} 
  &  {\bf Summation} \\[10pt]  

 &   {\bf on $R_N (z)$}  &  & \\  

1 Analytic  &  &  &   $\sum_{n=0}^{\infty}a_n z^n$  \\

 &  &  &    is convergent  \\

2 Singular,  &   $A \sigma ^{N} N! |z|^N $    &
  $B(t) =$   &  $g(z) = $  \\

opening angle $ = \pi $  & in ${\cal K}(\eta)$ and $N > N_{0} $ 
 & $ \sum_{n=0}^{\infty} \frac{a_n}{n!} t^n$ & $\frac{1}{z} 
\int_{0}^{\infty} {\rm e}^{-t/z} B(t){\rm d} t $ \\ \\

3 Singular,  &  $A \sigma ^{N} (N!)^{\rho} |z|^N$
  &   $B_{\rho}(t) =$  & $g_{\rho}(z) = $ 
\\

opening angle $ > 0 $  & in ${\cal L}(\rho)$, $N > N_{0}$ 
 & $ \sum_{0}^{\infty} \frac{a_{n}t^{n}}{\Gamma(n\rho + 1)}$
 & $\int_{0}^{\infty} {\rm e}^{-t}$ %{\rm exp}(-t^{1/\rho}) 
$B_{\rho}(t^{\rho}z) {\rm d}t$ \\ \\

4 Singular,  & $ A \mu (N) |z|^N$  & $M(t) =$  &  
$g_{m}(z)= \int_{0}^{\infty} M(tz) $  \\

opening angle $=0$ & in wedge ${\cal W}$, $N>N_{0}$ & $ 
\sum_{n=0}^{\infty} \frac{a_n}{\mu (n)} t^n $  & 
$ {\rm exp}(-{\rm e}^{t}) {\rm d} t$  \\ \\
\end{tabular}
\caption[]{{\bf Analyticity vs. high orders: a balance is 
needed for uniqueness}} 
\end{table}

A survey of typical cases is presented here in the Table 2, 
together with conditions for a unique determination of $f(z)$ from 
its asymptotic expansion. Conditions required for this are
placed under one and the same item. Of course, one does 
not expect that $f(z)$ will be
determined uniquely from its perturbation expansion in real
physical situations. Consequently, one cannot 
expect that the conditions placed in one item of 
Table 2 will be satisfied by a theory. This is indeed the 
case: QCD exhibits, according to \cite{'t Hooft} (see 
\cite{'t Khuri} for general proof), analyticity in the 
wedge ${\cal W}$\footnote{slightly modified by logarithmic 
terms} (see {\it item} 4 of the 1st column in Table 2), but the 
large-order behaviour of the series is governed by the $N!$ 
rule (see {\it item} 2 of the 2nd column). This is no surprise:
perturbation theory in QCD is incomplete and this discrepancy 
indicates the extent of non-uniqueness.     
The general rule for the use of Table 2 is as follows: 
To ascertain uniqueness in the determination of a function 
$f(z)$ out of its asymptotic series (\ref{as}), one has to 
check if the conditions in all columns in one row are satisfied 
for the case considered; otherwise the function is either 
not uniquely determined or may not exist.\footnote{The
integral representations for $g(z)$, $g_{\rho}(z)$ and
$g_{m}(z)$ placed in the last column of Table 2 exhibit a
symmetry which is revealed when appropriate substitutions
of integral variables are made.}

The disc ${\cal K}(\eta)$, the "drop" ${\cal L}(\rho)$ and the 
wedge ${\cal W}$ are depicted in Figs. 1 a, b, and c 
respectively.  

\begin{figure}
$$
\epsfxsize=\textwidth
\epsffile{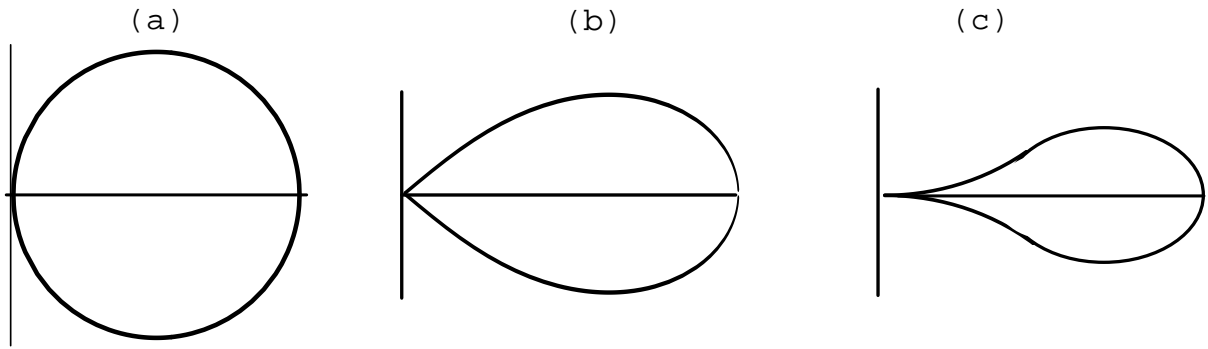}
$$
\caption[]{Three summation methods (see Table 2) for three 
different analyticity and boundedness domains: (a) the disc 
${\cal K}(\eta)$, (b) the drop ${\cal L}(\rho)$, and (c) the 
wedge ${\cal W}$. {\it Crucial is not the size of the domain, 
but the opening angle at the origin}.}
\end{figure}

Table 2 also shows that the large-order behaviour of the 
coefficients $a_{n}$ is not the only criterion of the (Borel 
or some other) summability of the series (\ref{as}). {\it 
Even a series with a very tame behaviour of the coefficients 
$a_n$ may not be Borel summable,} in spite of the radical 
suppression of the $a_{n}$ by the Borel factors $1/n!$. 
An example is discussed in \cite{MorozCJP}.

%%%%%%%%%%%%%%%%%%%%%%%%%%%%%%%%%%%%%%%%%%%%%%%%%%%%%%

\subsection{Useful facts on power series}

Certain important facts on power series are sometimes overlooked 
in physical considerations. It may therefore be useful to 
recall some of them here because spontaneous intuition is 
often misleading. I apologize for some repetition of the text.  

1. The divergence of a perturbation expansion does not 
signal an inconsistency or ambiguity in the theory. An 
example to illustrate this is the result of Ogievetsky 
\cite{Ogi}, who used the Borel summation method to sum a 
divergent perturbation series and obtained the exact result 
derived by Schwinger \cite{Schwi}, as is discussed here in 
subsection 1.1.   

The problem is not that of convergence or divergence, but 
whether the expansion uniquely determines the function or 
not. The method of Feynman diagrams allows one to find, at 
least in principle, all coefficients of the perturbation 
series, which may determine the function uniquely even if 
the series is divergent, and may not do so even if it is 
convergent. This depends on additional conditions.

2. The requirement (\ref{as}) of asymptoticity of a 
perturbation series
is not a formal assumption. It has physical content.

When a perturbation series is divergent, one often
re-interprets it automatically as an asymptotic series. This
assumption is weaker than that of convergence, but is not 
self-evident. It is not only technical either. Its physical 
meaning is that there is a very smooth transition between the 
system with interaction and the system without it.

For certain classes of observables the perturbation series 
is believed to be an asymptotic expansion.

3.  A very violent behaviour of the expansion coefficients $a_n$ 
at $n\rightarrow \infty$ might make us expect that no function 
with the property %$f(z) \sim \sum_{n=0}^{\infty}a_{n}z^{n}$ 
(\ref{as}) would exist. This
fear is not justified; it was proved by Borel and Carleman (see
\cite{Hardy} for details) that there are analytic functions 
corresponding to any asymptotic series.

4. Borel non-summability of a perturbation expansion alone 
does not signal an inconsistency or ambiguity in the theory. 
Borel summability (or its lack) is not a fundamental 
classification criterion (see, e.g., \cite{MorozCMP, Zhit}).

While large-order behaviour is often believed to be related to 
some very deep physics (see \cite{Zhit}), the Borel procedure 
is just one of many 
possible summation methods and need not be applicable always 
and everywhere. There are power series with mildly growing 
coefficients which are not Borel summable; the problem is to 
find a method (not necessarily the Borel one) which is 
appropriate for the case considered.  

Example of a non-Borel summable series (although composed of 
very tamely behaving coefficients) which, nevertheless, is 
summable by the method based on Theorem 1 is considered in 
\cite{MorozCJP} and discussed here in the end of subsection 
4.1, (\ref{mor1}). %, (\ref{mor2}).

5. If the $a_n$ behave very violently at $n \rightarrow \infty$ (so
that the Borel series $\sum_{n=0}^{\infty} a_{n} z^n/n!$ has 
zero convergence radius) one might expect that it would be 
sufficient to replace $n!$ in the denominator by a sequence 
${b_n}$ that grows faster than $n!$, in order to reach a more 
efficient suppression of the $a_{n}$. But this is done at a
price: stronger conditions on analyticity are required for the 
summation procedure to be unambiguous.

6. Asymptoticity is not a property of the series alone, but  
is contingent also upon the properties of the function expanded.
These properties (e.g., analyticity) must be examined
simultaneously with the asymptotic series, otherwise the same 
series can be summed to different functions.   

7. An asymptotic series need not be divergent. A wide-spread 
prejudice that gleams, often implicitly, in a number of papers 
is that an asymptotic series is always a divergent one. To 
see a quick counterexample, consider the function $F(z) 
= g(z) + A{\rm e}^{-\alpha /z}$ with $A$ real and 
$\alpha$ positive, where $g(z)$ is analytic at the origin.
This is a two-parametric class of functions $F(z)$
that all have, in the right half plane, the same asymptotic 
expansion, which is identical with the Taylor series for 
$g(z)$; so the asymptotic expansion of all the singular 
functions $F(z)$  is a convergent series because of analyticity 
of $g(z)$. This convergent series is asymptotic to many 
singular functions, %(most of which are singular at the origin), 
but its values converge only to one function, the analytic 
$g(z)$. They do so inside the Taylor circle, which extends 
up to the nearest-to-origin singularity of $g(z)$.

%%%%%%%%%%%%%%%%%%%%%%%%%%%%%%%%%%%%%%%%%%
%%%%%%%%%%%%%%%%%%%%%%%%%%%%%%%%%%%%%%%%%%%
\section{Operator-product expansion and analyticity}

%%%%%%%%%%%%%%%%%%%%%%%%%%%%%%%%%%%%%%%%

\subsection{Operator-product expansion}
A standard way to investigate the interplay of
perturbative and non-perturbative effects is the
operator-product expansion (OPE) \cite{Wilson, Zim}, in
which the singularities of an operator product are
expressed as a sum of non-singular operators with the
coefficients being {\it c}-number functions.
Being formulated before QCD, the operator-product expansion 
became a powerful (and actually the 
only) tool for examining non-perturbative effects in 
QCD. It consists in expanding the product of two 
local operators $A(x)$ and $B(y)$, for $(x-y)_{\mu}$ 
tending to zero, into a series of local operators 
${\cal O}_{n}(\frac{1}{2}(x+y))$,
\begin{equation}
A(x)B(y) \approx _{_{(x-y)_{_{\mu}} \rightarrow  0}}
\sum_{n} C_{n}(x-y) {\cal O}_{n}((x+y)/2).
\label{OPEx}
\end{equation} 
where the $C_{n}$ are {\it c}-number functions. The meaning 
of the symbol $\approx$ is discussed below. The  operators 
${\cal O}_{n}$  are ordered according to their canonical 
dimensions, and their quantum numbers must match those 
of $AB$. As a consequence, the coefficients $C_{n}$ are
ordered according to their decreasing singularity degree, 
and only a finite number of them are  
singular functions infinite at the origin, while the
remainder of the expansion is finite or tends to zero with 
vanishing distance. To any finite order in $x-y$, only a 
finite number of terms contribute. 

The short-distance behaviour of the
$C_{n}$ is obtained, by dimensional counting, from 
\begin{equation}
C_{n}(x) \rightarrow x^{d_{n}-d_{A}-d_{B}}(\ln
xm)^{p}(1+O(xm))
\label{asbe}
\end{equation}
with $d_{A}, d_{B}$ and $d_{n}$ being the dimensions of
$A, B$ and ${\cal O}_{n}$ respectively, and $m$ being a 
quantity of the dimension of mass. In this way the canonical 
dimension of ${\cal O}_{n}$ determines the singularity
degree of the corresponding $C_{n}$; the operators
${\cal O}_{n}$   with the smallest dimensions are
dominant at short distances, being multiplied with the
most singular coefficients $C_{n}$.

In momentum space, the expansion (\ref{OPEx})
takes the form (we put $y=0$)
\begin{equation}
{\rm i} \int {\rm d}x {\rm e}^{{\rm i}qx}A(x)B(0) \approx
\sum_{n} C_{n}(q){\cal O}_{n} .
\label{OPEq}
\end{equation} 

The symbol $\approx$ is understood differently in
different contexts. 

(i) While originally the validity of OPE was interpreted in 
the weak sense \cite{Wilson}, i.e., in that of sandwiching 
the product between two fixed states, from the very 
beginning it has been widely adopted \cite{Brandt} also to 
understand (\ref{OPEx}) and (\ref{OPEq}) as (formal) operator 
relations. 

(ii) Different is the interpretation also in the 
perturbative and in the non-perturbative context. The 
standard derivation of OPE
\cite{Wilson, Zim} relies on an analysis of Feynman
graphs and remains within the framework of
perturbation theory. The relations (\ref{OPEx}) and
(\ref{OPEq}) are often understood as asymptotic relations
(see, e.g., \cite {ItZ, Shif, VZNS}) which, in special
situations (free-field theory, \cite{Collins}), may 
become convergent Taylor series.

In application to non-perturbative phenomena, it is 
assumed \cite{SVZ} that the left- and the right-hand sides 
of (\ref{OPEq}) are approximately equal provided that a 
few first terms are taken. Then, 
at a critical dimension $d_{r}$, the series breaks down
and has to be supplemented with a term which is not of 
the $Q^{-2k}$ form. %indicated in (\ref{asbe}), (\ref{OPEq}). 
At higher values of $k$, other terms of 
different nature may appear. Relation (\ref{OPEq}) can 
be understood as an approximate equality with 
finite number of terms on the right-hand side \cite{SVZ}, 
with the hope that its validity and practical 
applicability (far from the point of expansion) 
can be established by phenomenological analysis of 
experimental data. Application of this approach to 
hadron resonances has been very successful; agreement 
with experimental data and internal consistency of the 
QCD sum rules scheme are impressive.

Non-perturbative effects manifest themselves also in 
the fact that, taking the vacuum-to-vacuum 
matrix element of (\ref{OPEq}), the terms on the 
right-hand side can be non-vanishing, whereas only the 
unit operator would survive within standard 
perturbation theory.

(iii) Various attitudes are adopted to the 
topical problem of the validity region (angle, rays) of
the expansion (\ref{OPEq}) in the complex $Q^2$ plane.
The expansion (\ref{OPEq}) reflects the properties of the
left-hand side at the point of expansion and in a {\it
subset} of its neighbourhood. 
In the euclidean ($Q^{2}>0$) region, the
expansion is well defined in the sense that the
separation of the large-distance contributions from the
short-distance ones is well defined \cite{Shif}. The
Minkowski ($Q^{2}<0$) region, on the other hand, is most 
important for applications, because it contains the 
spectrum of the physical states.  

This indicates that the expansion (\ref{OPEq}) can 
have different interpretations, according to the assumed 
physical content. These interpretations, however, have 
common mathematical features, which we are going to discuss 
in the next subsection. We adopt the approach (which is 
mostly used in phenomenology and is formulated, e.g., in 
\cite{ChibShif}) according to which the operator product 
expansion is originally built up in the euclidean region 
and then translated in the language of observables by 
analytic continuation  to the Minkowski region. (For 
problems of the definition of euclidean Fermi and Bose 
fields, we refer the reader to \cite{OstSchra} and 
references therein). 

Note that along directions in the complex plane
away from the euclidean ray different asymptotic
expansions may apply or may not exist at all.  

Predictions in the
Minkowski region are obtained by analytic continuation, 
which can be carried out under special assumptions 
of smoothness and stability. 
 
We shall now discuss conditions under which  
an asymptotic expansion can be extended to angles away 
from the euclidean semiaxis. It turns out that 
sign-definitness and integrability of the discontinuity 
along the cut in the Minkowski region play an important 
role.

%%%%%%%%%%%%%%%%%%%%%%%%%%%%%%%%
\subsection{Operator-product expansion away from
euclidean region}

The combination of the operator-product expansion with 
analyticity represents a solid basis for the calculation 
of the QCD observables. Such calculations require knowledge 
of the properties of the expansion in the complex energy 
plane away from the euclidean ray, and analyticity allows 
one to  establish them by means of analytic continuation. 
  
In practical applications of the operator-product
expansion one typically meets the following generic 
situation. Let $f(z)$ be holomorphic in $C$, the 
complex $z$ plane ($z=1/Q^2$), cut along the negative 
real semiaxis, $z\,\in \,[-\infty , 0]$. Let $f(z)$ be 
bounded by a constant,
\begin{equation}
|f(z)| < M
\label{V1}
\end{equation}
for $z\,\in \,C$ . Let the numbers $a_{k}$, $k=0,1,2,.
..n-1$, exist such that the following $n$ inequalities
\begin{equation}
|R_{k}(x)| < N_{k}x^k  
\label{V2}
\end{equation}
are satisfied for all $x={\rm Re} z$, $0<x<d$, with $d$ 
being a positive number, $n$ being a positive integer, and
\begin{equation} 
R_{n}(x) = f(x) - \sum_{k=0}^{n-1} a_{k}x^{k} . 
\label{V3}
\end{equation}
The problem is whether or in what sense the estimates
(\ref{V2}) apply also to rays away from the positive
real semiaxis. 

The condition (\ref{V1}) of boundedness by $M$ can be 
generalized to cases in which the function  of 
interest, $F(z)$, is bounded 
by a polynomial in $Q^{2}=1/z$ rather than by a constant; 
the theorem can then be applied either to $F(z)$
multiplied by the corresponding power of $z$, or to the
corresponding derivative of $F(z)$.  

Notice that our considerations in this section, being 
equally well applicable to the case of a fixed, finite value 
of $n$ as well as to the case of $n$ infinite, are not 
restricted to the case of asymptotic series. 
We find the case of a finite $n$ closer to physics because, 
first, one always knows only a finite number of the 
coefficients $a_{k}$  
and, second, theorems  dealing with a fixed value of $n$ 
are more transparent and their application to the case of
infinitely many inequalities (\ref{V2}) is straightforward. 

The solution to the problem very much depends on the 
character of the discontinuity of $f(z)$ on the cut.

%%%%%%%%%%%%%%%%%%%%%%%%%%%%%%%%%%%%%%
\subsubsection{The case of integrable discontinuity}

1. {\it Positive-definite discontinuity.} Let us first 
assume that $f(z)$ can be represented in the form
of a generalized Stieltjes integral \cite{BOrs},
\begin{equation}
f(z)=\int_{0}^{\infty}\frac{\rho(t)}{1+zt}{\rm d}t ,
\label{BOrs1} 
\end{equation}
where the discontinuity is positive-definite all along the 
cut, i.e., $\rho(t)$ is nonnegative for $t>0$. Let $\rho(t)$
be such that the integral in (\ref{BOrs1})
exists for $z$ complex, not negative, and that also the moments 
\begin{equation}
a_{k}=\int_{0}^{\infty} t^k \rho(t){\rm d}t
\label{BOrs2}
\end{equation}
exist for all nonnegative integers $k$. 

The function $f(z)$ has an asymptotic power expansion with
the coefficients $(-1)^k a_k$, 
\begin{equation}
f(z) \sim \sum_{k=0}^{\infty}(-1)^k a_k z^k
\label{BOrs3}
\end{equation}
for $z$ approaching zero from the positive side, $z
\rightarrow 0_{+}$. Indeed, 
\begin{equation}
R_{n}(z) = (-z)^{n}\int_{0}^{\infty}\frac{\rho(t)}{1+zt}t^{n} 
{\rm d}t
\label{BOrs4}
\end{equation}
and, consequently,
\begin{equation}
|R_{n}(z)|=\tilde{a}_{n}(z)|z|^n \leq a_{n}|z|^n  
\label{BOrs5}
\end{equation}
which, as expected, tends to zero one order faster than
$|z|^{n-1}$, the highest power subtracted from $f(z)$ (see  
(\ref{V3})). We use the notation 
\begin{equation}
\tilde{a}_{n}(z) =
\int_{0}^{\infty}\frac{\rho(t)}{|1+zt|}
t^{n}{\rm d}t .  
\label{BOrs6}
\end{equation}
Analogous estimates are obtained for $z$ approaching zero
along a ray in the complex plane (but away from the Minkowski
ray). %because $\tilde{a}_{n}(z)$ tends to $a_{n}$. 
In the right halfplane, certainly, $\tilde{a}_{n}(z) 
\leq a_{n}$ , but this is not the case for Re $z < 0$, 
where the general bound on $\tilde{a}_{n}(z)$ is worse, 
due to the presence of the cut in this halfplane. The bound  
can be obtained as follows. We have, 
denoting $z= r{\rm e}^{{\rm i}\phi}$, 
\begin{equation}
|1+zt| = (1+2rt \cos \phi + r^{2}t^{2})^{1/2} .
\label{Vr1}
\end{equation}
Considered as a function of $t$ at $t>0$ and $\phi$ fixed, 
$1/|1+zt|$ has its maximum at $t=t_{0}=-\frac{1}{r}\cos 
\phi$, where its value is $1/|\sin \phi|$. 
In this way we obtain (\ref{BOrs5}) and 

\begin{equation}
|R_{n}(z)| \leq \frac{1}{|\sin \phi|}\,a_{n}\, |z|^n 
\label{Vr3} 
\end{equation}
for Re$\,z > 0$ and Re\,$z < 0$ respectively. Comparing
(\ref{Vr3}) with (\ref{BOrs5}), we see how the factor 
$1/|\sin \phi|$ makes the estimate looser when the ray
approaches the direction of the cut, i.e., when $\phi$ 
approaches zero.

Possible applications to some QCD processes are discussed in 
subsection 5.2.

%%%%%%%%%%%%%%%%%%%%%%%%%%%%%%%%%%%%%%%%%%%%

2. {\it Sign-indefinite discontinuity.} The estimates are 
worse if the discontinuity along the 
cut is not positive definite. Physical relevance of such 
situations is pointed out in \cite{Shif}. If the
integral representation (\ref{BOrs1}) is still
admitted,  be it with $\rho(t)$ sign-indefinite or even
a complex-valued function on the integration interval,
the resulting bounds (\ref{BOrs5}) and (\ref{Vr3}) 
have to be modified in the following
sense. The function $\rho(t)$ is assumed to be such that 
the integral (\ref{BOrs1}) is absolutely convergent and 
that the moments  of $|\rho(t)|$ , 
\begin{equation}
\bar{a}_{k}=\int_{0}^{\infty}t^{k}|\rho(t)|{\rm d}t ,
\label{BOrs7}
\end{equation}
are convergent integrals; this excludes  wild 
oscillations  of the discontinuity along the cut. The
estimates  (\ref{BOrs5}) and (\ref{Vr3}) 
are replaced by 
\begin{equation}
|R_{n}(z)| \leq  \bar{a}_{n} \, |z|^n 
\label{Vr22} 
\end{equation}
and
\begin{equation}
|R_{n}(z)| \leq \frac{1}{|\sin \phi|}\,\bar{a}_{n}\, |z|^n 
\label{Vr33} 
\end{equation}
respectively. This may imply a considerable deterioration 
of the estimate relative to a sign-definite $\rho(t)$: if 
the discontinuity on the cut oscillates, $\bar{a}_{n}$ and
$|R_{n}(z)|$ will have very little to do with the expansion 
coefficient $a_{n}$.  

Better estimates on $|R_{n}(z)|$ than (\ref{BOrs5}), 
(\ref{Vr3}) and, respectively, (\ref{Vr22}), 
(\ref{Vr33}) cannot be obtained unless some special
assumptions are made about the explicit form of the
function $\rho(t)$ in the respective cases. 

%%%%%%%%%%%%%%%%%%%%%%%%%%%%%%%%
\subsubsection{General case} 

It is interesting to discuss a further generalization,
although its relevance for physics may still be unclear.  
In the general case, when the integral (\ref{BOrs1}) is not
absolutely convergent or the moments (\ref{BOrs7}) are not
defined, we can use a theorem by I. Vrko\v{c}  
(see \cite{Vrkoc}) to obtain an estimate. Assumed are 
holomorphy of $f(z)$ and the bound (\ref{V1}) on $f(z)$ 
to be valid in an angle $2 \alpha$ bisected by the real 
axis, and the inequality 
\begin{equation}
|R_{n}(x)| < N_{n}x^n  
\label{V22}
\end{equation}
to hold for a positive integer $n$ and for $0<x<d$. When   
applied to $\alpha = \pi$ (the values of $n$ and $d$
being fixed), the theorem implies that, along all rays
between $-\pi(1-2^{-l})$ and $\pi(1-2^{l})$, $R_{m}(z)$ 
satisfies the following inequality:
\begin{equation}
|R_{m}(z)|<\hat N_{m}|z|^{n/2^{l}}
\label{V4}
\end{equation} 
for all $|z|<d_0$, where $d_{0}$ is a positive number 
and $l$ is a nonnegative integer. The coefficient 
$\hat N_{m}$ in front of $|z|^{n/2^l}$ is a linear 
combination of the $|a_{k}|$ for $k$ running from     
$m=[(n-1)/2^{l}]+1$ to $n-1$, $[q]$ denoting the 
integer part of the real number $q$. 

This result is of relevance for 
expansions both with a finite and with infinite number 
of terms. Note that the bound (\ref{V22}) on $R_{n}(x)$ 
for $x$ positive implies, for $z$ complex, the estimate 
(\ref{V4}) on $R_{m}(z)$, where $m$ is smaller than $n$. 
This resulting bound for complex $z$ is considerably looser
than that for $x$ real, the number of exploitable terms 
being reduced from $n$ down to $m=[(n-1)/2^{l}]+1$. 
This effect is the price paid for the extended validity 
region of (\ref{V4}), and is becoming more and more 
pronounced with increasing $l$, i.e., with increasing 
angular deflection from the euclidean interval. We also 
see that the higher-order expansion coefficients $a_{k}$ 
(for $k=m+1,.., n-1$) cannot 
be used to improve the bound  along  a  complex ray. 

The result shows quantitatively how quickly the  
expansion breaks  down when the angle becomes large, i.e., 
when one approaches the cut on the Minkowski side of the 
real axis. 
As is usual in analytic extrapolations \cite{CiuCutFiPi}, 
a point-by-point continuation from approximate data is 
impossible, and an extrapolation up to the cut can be 
safely done onto an interval but not to a point. The 
procedure should be based on statistical 
analysis of experimental data,  \cite{CutPi}. 

Since the estimate (\ref{V4}) might seem rather loose (note 
that $n$ in the exponent on the right hand side is divided 
by $2^l$), it is interesting to look for a function which 
would saturate it. A set of functions saturating 
(\ref{V4}) for different nonnegative integers $n$ can
be generated by using the function  
$\exp(-{\rm i} a(\ln z)^2)$ with $a$ 
real.\footnote{I am grateful to I. 
Vrko\v{c} for pointing out this to me.} 

Further modifications of the results of this section 
will arise when the logarithmic 
dependence of the expansion coefficients $C_{n}(x)$ 
indicated in (\ref{asbe}) is taken into account. 
Physical relevance of this generalization is evident.

%%%%%%%%%%%%%%%%%%%%%%%%%%%%%%%%%
\section{Renormalons and instantons}
%%%%%%%%%%%%%%%%%%%%%%%%%%%%%
\subsection{Singularities in the Borel plane}

The integral (\ref{BS}) defines the Borel sum of the series (\ref{Pt2}) 
if the conditions 1), 2) and 3) from subsection 2.1 are satisfied.

The condition 1) would be violated if the $a_{n}$ were to
grow faster than $n!$. As follows from Table 1, this is not 
the case in typical situations.

We generally do not know the nature or distribution of 
singularities 
to assess the validity of the condition 2).  Analyses
of large-order behaviour indicate that 2) is violated 
by an infinite set of singularities located  on the real 
axis (renormalons, instantons). Then, the integral 
(\ref{BS}) is not well-defined and should not be used.  
In quantum theory, however, it is nevertheless used, 
with the aim to reduce the problem of summation    
to that of defining the integral (\ref{BS}) by means of 
a suitable modification of the integration contour. 

The spectrum of ambiguitites is wider, however. By 
introducing the renormalon language we 
consider only special sets of equidistant singularities 
(poles, branch points)   
whose  parameters have physical interpretation. 
Those lying on the positive real semiaxis produce 
ambiguity and remind us that non-perturbative 
input has to be added.

The electromagnetic current--current correlation function is a useful
example to discuss a typical structure of singularities in the complex
 coupling-constant plane. Denoting this function $\Pi ^{\mu \nu}$,
\begin{eqnarray}
\Pi ^{\mu \nu} = {\rm i} \int {\rm d}^4 x {\rm e}^{-{\rm i}qx} 
\langle 0 | \, {\rm T} (j^{\mu}(x) j^{\nu}(0)) \, | 0 \rangle \\ 
=(g^{\mu \nu}q^2 - q^{\mu}q^{\nu}) \Pi (-q^2)
\end{eqnarray}
and taking $R$, the ratio of the total cross section for ${\rm e}^+
{\rm e}^- \rightarrow {\rm hadrons}$ to that for ${\rm e}^+ {\rm e}^-
\rightarrow \mu^{+}\mu^{-}$, which is related to the imaginary
part of $\Pi$, 
\begin{equation}
R(s) = 12 \pi {\rm Im} \, \Pi (s+{\rm i}0^{+}) \, ,
\label{Im}
\end{equation}
we introduce a modified quantity 
$D(Q^2)$ defined as
\begin{equation}
D(Q^2 ) = - 4\pi ^2 Q^2 
\left(\frac{{\rm d}}{{\rm d}Q^2}\right) \Pi (Q^2 ) 
\label{D} 
\end{equation}
(where $Q^{2} = - q^{2}$, i.e., $Q^2 >0$ in the euclidean 
domain), to avoid inessential logarithmic 
terms and the dependence on the renormalization 
scale $\mu$ \cite{BYZhai}). The perturbation expansion of $D$ in 
powers of the coupling constant has the form
\begin{equation} 
D \sim 1+\frac{\alpha_{s}(Q^2)}{\pi} \sum_{n=0}^{\infty} 
D_{n} (\alpha_{s}(Q^2))^n.
\end{equation}

The dependence of $D$ on $\alpha_{s}(Q^2)$ exhibits a
complex structure of singularities in the coupling constant 
plane at and around the origin \cite{'t Hooft}. Their 
nature can be conveniently displayed by studying the 
corresponding Borel transform. The behaviour of the running 
coupling constant $g(\mu^2)$ is determined by the $\beta$ 
function, $\beta(g^{2}(\mu^2))\equiv \mu^2 ({\rm
d}/{\rm d}\mu^2) g^{2}(\mu^2)$, which has the perturbative 
expansion $\beta(g^{2}) \sim -b_{0}g^{4}-b_{1}g^{6}-...\,$. 

The structure of the singularities of the Borel transform 
that are nearest to the origin can be seen from the 
following (simplified, see \cite{Zakh, BSUV}) expression for 
$D$
\begin{equation}
Q^2 \int{\rm d}k^2 \frac{k^2 \alpha_{s}(k^2)}{(k^{2}+Q^{2})^{3}} , 
\label{renor}
\end{equation}
which at the infrared and the ultraviolet end of the spectrum has
the following behaviour
\begin{equation}
\alpha_{s}(Q^2) \sum_{n}[b \alpha_{s}(Q^2)/(4\pi)]^n n!/2^{n+1}
\label{IR}
\end{equation}
and
\begin{equation}
\alpha_{s}(Q^2) \sum_{n}[-b \alpha_{s}(Q^2)/(4\pi)]^n n!
\label{UV}
\end{equation}
respectively, where $b$ is the first coefficient in the
Gell-Mann--Low function. This corresponds to an $n!$
large-order behaviour, with the nearest singularities located in
the Borel plane at $8\pi/b$ and $-4\pi/b$ respectively.  

Perturbation theory suggests the following structure of the 
singularities of the Borel transform \cite{BYZhai,Zakh}: 

(1) {\em Instanton--anti-instanton pairs} \cite {Lipatov,Brezin}
generate equidistant singularities along the positive real axis
starting at $t = 4$, for $t = 4l, l=1,2,...$. Balitsky \cite{Balitsky}
calculated the behaviour of $R_{{\rm e}^+{\rm e}^- \rightarrow 
{\rm hadrons}}$ near $t=4$ and found the leading ${\rm I - 
{\bar I}}$ singularity to be a branch point with the power 
$\frac{11}{6}(N_{\rm f}-N)$, where $N$ and $N_{{\rm f}}$ is 
the number of colours and of flavours respectively.

(2) {\em Ultraviolet renormalons} are generated by contributions
behaving as $c_{k+1} \sim (-b_0 /l)^k k!$ for $l=1,2,...$, leading to
singularities located at $t = - 2 l/b$ \, on the negative real axis
(where $b=8 \pi^2 b_{0}$), with a branch point for $l=1$.

(3) {\em Infrared renormalons} are generated by contributions behaving
as $c_{k+1} \sim (b_0 /l)^k k!$, $l=2,3,...$, leading to singularities 
located at $t=2l/b $. Near the first of the points, the singularity 
behaves \cite{Mueller} as $(bt - 4)^{-1-2\lambda/b_{0}}$, 
$\lambda = b_{1}/b_{0}$. By absence of a dimension-two  
condensate the existence of the $l=1$-singularity 
is excluded.

Other singularities are not assumed to exist in the Borel plane. 
But there may be still another source of ambiguities, due to
possible singularities at infinity. They may be non-Borel 
summable, no matter how tame the large-order 
growth may be! A trivial example is the
series $\sum_{n=0}^{\infty} z^n$, whose coefficients are
constant. It has no singularities in the Borel plane,
but the series is Borel summable only for Re$z<1$, because
otherwise the Borel integral does not exist. 
A more sophisticated example is $\sum_{n=0}^{\infty} 
c_{n}z^{n}$ with
\begin{eqnarray}
c_{n}=c^{n} \mu (n) \nonumber \\
\mu(n) = \int_{0}^{\infty} {\rm exp(-e}^{t})\,\, t^{n} {\rm d}t 
\label{mor1} \\
c=|c|{\rm e}^{{\rm i}\theta},\,\,\,\,\, 0 < \theta < \pi /2  
\nonumber 
\end{eqnarray}
(see ref. \cite{MorozCJP}, Example 3). 
Note the double exponential in the integrand, which makes $\mu (n)$
and $c_{n}$ rise extremely slowly. At first sight, this series 
should be Borel summable, even more that there are no singularities 
in the Borel plane, the convergence radius (condition 2)) being 
infinity. But the condition 3) is not satisfied: the integral 
(\ref{BS}) does not exist. Indeed, 
\begin{equation}
\int_{0}^{\infty}{\rm d}t {\rm e}^{-t/z}
\sum_{n=0}^{\infty}\frac{c_{n}}{n!}t^{n}
\label{mor2}
\end{equation}
is divergent, because $B(t)$ grows faster than ${\rm e}^{t}$ with $t
\rightarrow +\infty$. But the series $\sum_{n=0}^{\infty} c_{n}z^{n}$
is summable by Moroz's method \cite{MorozCMP}.

%%%%%%%%%%%%%%%%%%%%%%%%%%
\subsection{A further generalization of  Borel transformation}
%%%%%%%%%%%%%%%%%%%%%%%%%%

The functions $B_{\rho}(t)$ and $M(t)$ defined in Table 2 are
generalizations of the Borel transform, which can be used in the
various situations listed in Table 1 to reduce non-uniqueness, 
provided some additional information is available. More about 
the properties of $B_{\rho}(t)$ and $M(t)$ can be found in
\cite{MorozCMP,MorozCJP,MorozOTH,Fischer} and in references 
therein.

I will now discuss another type of generalization of the notion 
of Borel transform, \cite{BrownYaffe,BYZhai,Beneke1}, which 
makes use of the specific structures mentioned in the previous
subsection.  

Let me first make a general remark. After having exposed
mathematical methods in sections 1 and 2, we have passed to
practical aspects of the operator-product expansion (estimate
of the remainder after the $n$-th term, see section 3), and of 
the assumption (in section 4) 
that two-point Green's functions have special singularities  
in the Borel plane, the instantons and the renormalons, a 
structure that is now almost universally adopted. Existence of 
these structures in the Borel plane is based on various
mathematical models developed in late 70's and early 80's, but 
nowadays they are mostly considered as true features of Nature. 

Brown, Yaffe and Zhai and Beneke \cite{BrownYaffe,BYZhai,Beneke1} 
use information about the structure of the first infrared renormalon.
Assuming first the special case of a one-term $\beta$ function, 
\begin{equation}
\beta (g^{2}) = -b_{0}\,g^{4} \, ,
\label{beta}
\end{equation}
expanding ${\rm Im} \, \Pi$ and $\Pi$ in powers of the coupling constant
with the expansion coefficients $a_{n}$ and $c_{n}$ respectively  
and defining their respective Borel transforms
\begin{equation}
A(t) = \sum_{n=1}^{\infty} \frac{n a_{n}}{\Gamma(n+1)}\, t^n
\end{equation}
and
\begin{equation}
C(t) = {\tilde c_{0}} + \sum_{n=1}^{\infty} \frac{n c_{n}}
{\Gamma(n+1)}\, t^n \,,
\end{equation}
Brown and Yaffe \cite{BrownYaffe} obtain, by comparing the expansion 
coefficients, the following relation between $A$ and $C$:
\begin{equation}
A(t) =  \sin (b_0 t) C(t)  \,   ,  
\label{AC} 
\end{equation}
which turns out to be a consequence of renormalization-group
invariance \cite{Beneke1}. Defining a modified Borel transform
${\cal F}(t)$ by
\begin{equation}
{\cal F}(t) = \sum_{n=0}^{\infty} \frac{\Gamma(1+\lambda t)}{
\Gamma(n+1+\lambda t)} f_{n}t^{n}     
\label {F}
\end{equation}
(thereby accounting for the first infrared renormalon), the authors of
\cite{BYZhai,Beneke1} consider the case of a general beta
function, chosen in such a scheme that its inverse contains
two terms:
\begin{equation}
1/\beta(g^{2}) = -1/(b_{0}g^{4}) + \lambda /(b_{0}g^2 ) .
\end{equation}
They find that, for this form of $\beta(g^2)$, the relation 
(\ref{AC}) remains valid also for ${\cal A}(t)$ and ${\cal C}(t)$, 
the modified (according to (\ref{F})) Borel transforms of 
${\rm Im} \, \Pi$ and $\Pi$ respectively:
\begin{equation}
{\cal A}(t) =  \sin (b_0 t) \, {\cal C}(t)  \,   .
\label{ACZhai} 
\end{equation}
In this way, the endeavour to retain the relation (\ref{AC}) also for 
the case of a general $\beta$ function motivates the authors to 
introduce the generalized Borel transform in the form (\ref{F}). 
  
It was already pointed out that the concept of renormalon is at 
present applied to concrete physical situations. This poses a  
topical problem: To what extent are renormalons physical 
concepts and to what extent are they just artefacts of our, 
maybe inadequate, formalism? Some authors \cite{Krasnikov} argue 
that the series of renormalon-type graphs is ill-defined, or that
infrared renormalons might be an attribute of the renormalization
scheme used. The Borel plane formalism is, nevertheless, 
successful in phenomenology.  

%%%%%%%%%%%%%%%%%%%%%%%%%%%
\subsection{Phenomenology in the Borel plane}

Soper and Surguladze \cite{Soper} use 
information on the singularities in the Borel plane  
to compute the ratio $R$ of the width for
Z$\rightarrow$ hadrons to that for Z$\rightarrow$
e$^{+}$e$^{-}$. The result is impressive.   
Considering $R$ as a function $R(s)$ of $s$, the c.m. 
energy squared of the  e$^{+}$e$^{-}$ annihilation, one 
obtains the measured $R$ as $R(M^{2}_{{\rm Z}})$. Representing 
$R(s)$ and $D(s)$ as 
$R_{0}[1+{\cal R}(s)]$ and $D_{0}[1+{\cal D}(s)]$ 
respectively with ${\cal R}$  and  ${\cal D}$ expanded in 
powers of the coupling constant $\alpha_{s}(s)/\pi$,  
the authors of \cite{Soper} use, in view of the
relations (\ref{Im}) and (\ref{D}), the formula
\begin{equation}
{\cal R}(s) = \frac{1}{2\pi}\int_{-\pi}^{\pi}{\rm d}\theta{\cal
D}(s{\rm e}^{{\rm i}\theta})
\label{Sop1}
\end{equation}
to calculate ${\cal R}(s)$ from ${\cal D}(z)$. Both ${\cal R}$
and ${\cal D}$ are expanded in perturbation series to third 
order, where the coefficients are explicitly known.  
Assuming now a standard 
large-order behaviour of the perturbative expansion of 
${\cal D}$, the authors %of \cite{Soper} 
write the Borel representation for it,
\begin{equation}
{\cal D}(Q^2)=\int_{0}^{\infty}{\rm d}z \exp(-\pi
z/(\alpha_{s}(Q^2))\tilde{\cal D}(z),
\label{Sop2}
\end{equation}
where the Borel-transform coefficients $\tilde{\cal D}_{n}$ 
have better large-order behaviour than the ${\cal D}_{n}$, 
those of the original series.   
Combining (\ref{Sop1}) with (\ref{Sop2}),
one expresses ${\cal R}$ directly in terms of $\tilde{\cal 
D}(z)$ through an integral over $z$ from 0 to $\infty$, in which 
the upper limit can be safely replaced by $z_{{\rm max}}$, with
$\beta_{0}z_{{\em max}}$ around 0.5 or higher. The ultraviolet 
renormalon at  $\beta_{0}z=-1$, being the singularity closest 
to the origin, controls the large-order behaviour of the 
perturbative series; the authors of \cite{Soper} move it farther
by means of the conformal mapping
\begin{equation}
\beta_{0}(\zeta)=4\frac{\sqrt{1+\beta_{0}z}-1}
{\sqrt{1+\beta_{0}z}+1}\,\, ,
\label{KZ}
\end{equation}
thereby improving the convergence rate of the power series in
the Borel plane. Note that the other set of dangerous
singularities, the infrared renormalons (the nearest of which 
is at $\beta_{0}z=+2$), are not pushed away from the origin by 
(\ref{KZ}), but on the contrary become closer to it. Although 
the infrared renormalons  
could be pushed away by means of the optimal 
conformal mapping \cite{CF} (which leads to the fastest 
convergence), Soper and Surguladze \cite{Soper} prefer to 
soften  
the nearest infrared renormalon by multiplying $\tilde{\cal D}$ 
with its power,  because they know the numerical value of the 
exponent. Making good use of this valuable 
piece of information, 
they reach a high accuracy in determining
$\alpha_{s}(M^{2}_{{\rm Z}})$, estimating that the 
uncertainty arising from QCD is less than  0.5 per cent.

%%%%%%%%%%%%%%%%%%%%%%%%%%%%%%%%%
\subsection{Renormalons and OPE}

A scheme to use renormalons to construct non-perturbative 
terms in the operator-product expansion has been proposed by 
Grunberg \cite{Grun}. Defining from $\Pi(Q^2)$ the
renormalization-group invariant quantity 
\begin{equation}
R(Q^2)=\frac{{\rm d}\Pi}{{\rm d}\ln Q^2}- 
\frac{{\rm d}\Pi}{{\rm d}\ln Q^2}|_{\alpha=0} ,
\label{Grun1}
\end{equation}
he uses the OPE representation
\begin{equation}
R(Q^2)=R_{{\rm PT}}(\alpha)+G_{0}(\alpha) + ... ,
\label{Grun2}
\end{equation}
where $\alpha=\alpha(\mu)$ is the coupling at scale $\mu$,
$R_{{\rm PT}}(\alpha)=\sum_{n=0}^{\infty} r_{n}\alpha^{n+1}$ the
perturbative contribution and $G_{0}(\alpha)$ is the leading
condensate contribution. For $G_{0}(\alpha)$, the renormalization 
group equation yields
\begin{equation}
G_{0}(\alpha)=C(\mu^2/Q^2)^{d/2}\exp(d/(2\beta_{0}\alpha))\alpha
^{\delta}[1+O(\alpha)] ,
\label{Grun3}
\end{equation}
where $\delta$ is expressed through the $\beta$-function
coefficients, and the $\mu$-dependence of the factor 
$(\mu^2/Q^2)^{d/2}$ is compensated by that of
the subsequent exponential function. 

The function $R_{{\rm PT}}(\alpha)$ is then approximated by first
infrared renormalon singularity. Using  the Borel
representation 
\begin{equation}
R_{{\rm PT}}(\alpha)=\int_{0}^{\infty}{\rm d}z
\exp(-z/\alpha)R^{\rm B}_{\rm PT}(z)
\label{Grun4}
\end{equation}
with $R^{{\rm B}}_{{\rm PT}}(z)$ approximated by the first
renormalon  $R^{{\rm B}}_{{\rm PT,0}}(z)$,
\begin{equation}
R^{{\rm B}}_{{\rm PT,0}}(z)=K(\mu^{2}/Q^2)^{-z_{0}\beta_{0}}
(1-z/z_{0})^{-\gamma}[1+O(1-z/z_{0}] ,
\label{Grun5}
\end{equation}
where $K$ is a scale-independent factor, Grunberg arrives 
\cite{Grun} at
\begin{equation}
{\rm Im}R_{{\rm PT,0}}(\alpha)=\pm Kz_{0}^{\gamma}
\Gamma(1-\gamma)\sin(\pi(1-\gamma))(\mu^{2}/Q^{2})^{-z_{0}
\beta_{0}}\exp(-z_{0}/\alpha)\alpha^{1-\gamma}[1+O(\alpha)] .
\label{Grun6}
\end{equation}
The sign depends on the choice of the integration contour; thus,
the resummed $R_{{\rm PT}}(\alpha)$ cannot determine $R$ uniquely.
Consistency requires the ambiguity to be removed by some
non-perturbative input. Assuming $C$ in (\ref{Grun3}) complex,
Grunberg \cite{Grun} compares (\ref{Grun6}) with (\ref{Grun3}) to find 
relations connecting $z_{0}\beta_{0}$ and $\gamma$ with $d$ and 
$\delta$ and to fix the position and the power of the singularity. 

This method allows non-perturbative terms in the operator-product 
expansion to be constructed from perturbation theory.  
The author gives a convergent scheme to deal with perturbation
theory in presence of infrared renormalons. The procedure can be
extended to non-leading IR renormalons. 

Grunberg's result shows that there are ``non-perturbative'' 
terms in the operator-product expansion which have their origin 
in perturbation theory. The author considers also the possibility 
that all ``non-perturbative terms'' in the expansion, taken to 
all orders, could be determined in principle from perturbation 
theory. This interesting possibility could be a good starting point 
for approximations, although it does not mean 
that all QCD could be fully  
described in terms of perturbative concepts. Within 
perturbation theory one can account for the infrared domain 
up to terms of the form  
\begin{equation}
\exp(m/(\beta_{0}\alpha(Q)) \sim (\Lambda^{2}/Q^{2})^m 
\label{BB}
\end{equation} 
(see a discussion in \cite{BB1}), where $m$ is the order of IR 
renormalon. 
On the other hand, an additional non-perturbative input may also 
have implications in the Borel plane. %In general, t
The presence of 
renormalons can signal deficiency of the perturbative 
approach and indicate a nonperturbative supplement.

%%%%%%%%%%%%%%%%%%%%%%%%%%%%%%%%%

%%%%%%%%%%%%%%%%%%%%%%%%%%%%%%%%%

\section{Further applications}

%%%%%%%%%%%%%%%%%%%%%%%%

\subsection{Resummation of bubble chains}

The renewed interest in calculating higher-order perturbative
corrections and in examining the high-order behaviour of perturbative
series is intimately related to the investigation of renormalization
scale and scheme dependence of a truncated series as well as to
attempts to estimate its uncalculated remainder. In the past various
criteria for finding a suitable renormalization prescription were
proposed; they are based on an estimate of the size of the remainder.
Examples are Stevenson's principle of minimal sensitivity \cite{PMS},
Grunberg's notion of effective charge \cite{GEF},  and the BLM method 
of scale setting \cite{BLM} by Brodsky, Lepage and Mackenzie. The 
BLM prescription, which is based on an analogy with QED, was further 
developed recently. It is a method of estimating higher-order
perturbative corrections of a physical quantity, provided that the
first approximation is known. It consists in the use of some ``average 
virtuality'' as scale in the running coupling. Instead of working with
fixed scale, 
\begin{equation}
\alpha_{s}(Q^2 ) \int {\rm d}^{4}k F(k, Q)\, ,
\end{equation}
one averages over the logarithm of the gluon momentum:
\begin{equation}
\alpha_{s}(Q_{BLM}^{2} ) \int {\rm d}^{4}k F(k, Q) \equiv 
\alpha_{s}(Q^{2}) \int {\rm d}^4 k \left(1-\frac{\beta_{0}}{4\pi}
\alpha_{s}(Q^2)\ln \frac{-k^2}{Q^2}\right) F(k,Q) \,.\,\,\, \label{*}
\end{equation}
This replacement amounts to accounting for higher-order terms in 
powers of $\alpha_{s}(Q^2)$ by making use of the 
renormalization-group evolution
\begin{equation}
\alpha_{s}(-k^2 )= \alpha_{s}(Q^{2}) \sum_{n=1}^{\infty}
\left(\frac{\beta _{0} \alpha_{s}(Q^{2})}{4\pi }\right)^{n-1} 
(-\ln (-k^2/Q^2))^{n-1} 
\label{sum}
\end{equation}
(in the leading-logarithm approximation) and retaining only the 
first two terms in the sum. This approach 
was generalized \cite{BenekeBraun,BBB1,Neubert} by introducing
the running coupling constant $\alpha_{s}(k^{2})$ directly into 
the vertices of Feynman diagrams, with $k$ being the momentum 
``flowing'' through the line of the virtual gluon. This modification 
means replacement of (\ref{*}) by
\begin{equation}
\alpha_{s}(Q^{*2}) \int {\rm d}^{4}k F(k, Q) =   \int {\rm d}^4 k 
\alpha_{s}(-k^2) F(k,Q) \,\,\, \label{**}
\end{equation}
with $\alpha_{s}(x^2)=4\pi/(\beta_{0}\ln(x^2 /\Lambda^{2}_
{QCD})$. Note, however, that the beta function 
$\beta(\alpha_{s}(Q^{2}))$ is approximated by its first term only.

This method has been applied in phenomenology to various QCD 
observables, such as $\tau$ decay hadronic width and heavy-quark 
pole mass \cite{BBr,Neubert}, semileptonic B-meson decay 
\cite{BBB2} and the Drell--Yan process \cite{BBr}. It makes 
maximal use of the information contained in the one-loop 
perturbative corrections combined with the one-loop running of the 
effective coupling, thereby providing a natural extension of the BLM 
scale-fixing prescription. 

Methods generalizing the scale-setting procedure developed by 
Brodsky, Lepage and Mackenzie met with criticism \cite{Chyla,LTM}   
showing that the summation based on ``naive 
non-abelization'' is re\-nor\-ma\-lization-pre\-scrip\-tion 
dependent. Further research will clarify the issue; some of 
its aspects are discussed in Sec. 6.  

A resummation
that is renormalization-scheme invariant in the large-$N_{f}$
approximation has recently been proposed by Maxwell and 
Tonge \cite{MT}. It is based on approximating the effective
charge beta-function coefficients by %retaining 
the part produced by the highest power of 
$b=(11N_{c}-2N_{f})/6$, and accounts for exact all-orders 
results in this  approximation. This method allows one to
include, in any renormalization scheme, the exact perturbative 
coefficients  not only at NLO but also at NNLO. The
approach is used to critically examine the reliability of 
calculations of fixed-order perturbation theory for various 
observables and sum rules. Application of
the RS-invariant resummation to different physical processes
indicates varying reliability of the standard fixed-order
perturbation theory.  

Ellis et al. \cite{Ellis} use Pad\'e approximants to develop 
another method of resumming the QCD perturbative series. The 
authors test their method on various known QCD results and 
find that it works very well; in particular, they reach a 
weaker renormalization-scale dependence of the calculations. The 
problem why Pad\'e approximants reduce the renormalization-scale
dependence has recently been considered in ref. \cite{Gardi}. 
The author finds that in the ``large $\beta_{0}$ limit'' (when the 
beta function is dominated by the 1-loop contribution) diagonal 
Pad\'e approximants become renormalization-scale independent, 
thereby sharing an important feature of the exact result.

%%%%%%%%%%%%%%%%%%%%%%%%%%%%%%%%%%%%%%%%

\subsection{Use of operator-product expansion}

%%%%%%%%%%%%%%%%%%%%%%%%%%%%%%%%%%%%%%%%
Operator-product expansion has proved to be very useful in
practical calculations %, in particular in the investigation of 
in connection with the tau
decay hadronic width, the heavy quark pole mass, heavy-light 
quark systems, semileptonic B-meson decay, heavy quarkonia 
and the Drell-Yan process. 
Renewed interest in the evaluation of power corrections to
QCD predictions has resulted in attempts to understand the
implications of the presence of renormalon singularities  
in these processes.        

Starting from analyticity and operator product expansion 
Braaten, Narison and Pich \cite{BNP} used the 
measurements of the $\tau$ decay rates to determine the 
QCD running coupling constant at the scale of the 
$\tau$ mass $M_{\tau}$=1.784 GeV. The ratio $R_{\tau}$,
\begin{equation}
R_{\tau} = \frac{\Gamma[\tau^{-} \rightarrow \nu_{\tau}
{\rm hadrons}(\gamma)]}{\Gamma[\tau^{-} \rightarrow 
\nu_{\tau}{\rm e}^{-}\bar{\nu}_{\rm e}(\gamma)]}
\label{Rta1}
\end{equation}
(where $(\gamma)$ represents possible additional photons or 
lepton pairs), is represented in the form
\begin{equation}
12 \pi \int_{0}^{M_{\tau}^2} (1-s/M_{\tau}^2)^2 
(1+2s/M_{\tau}^2) {\rm Im}\Pi(s)\frac{{\rm d}s}{M_{\tau}^2}
\label{Rta2}
\end{equation} 
where $\Pi(s)$ is a  combination of the two-point 
correlation functions for the vector $V_{ij}^{\mu}=
\bar{\psi}_{j}\gamma^{\mu}\psi_{i}$ and axial vector 
$A_{ij}^{\mu}=\bar{\psi}_{j}\gamma^{\mu}\gamma_{5}\psi_{i}$ 
colour singlet quark currents, with coefficients given by 
the elements $V_{{\rm ud}}$ and $V_{{\rm us}}$ of the 
Kobayashi-Maskawa matrix, the subscripts $i, j=$u,d,s 
denoting light quark flavours. 

This integral cannot at present be calculated from QCD, because
the hadronic functions are sensitive to the non-perturbative
effects that confine quarks in hadrons. But one can make use of
the properties of the correlating functions, which are
known to be analytic in the complex $s$-plane cut along the
positive real semiaxis. This allows the integral in (\ref{Rta2})
to be expressed as a contour integral along the circle of radius
$M_{\tau}^{2}$:
\begin{equation}
6 \pi {\rm i} \oint_{|s|=M_{\tau}^2} (1-s/M_{\tau}^2)^2 
(1+2s/M_{\tau}^2) \Pi(s)\frac{{\rm d}s}{M_{\tau}^2} \, .
\label{Rtau}
\end{equation}
Representing now $\Pi(s)$ as the operator product expansion 
over local gauge invariant operators, the authors of \cite{BNP} 
point out that the double zero in the kinematic factor 
of the integrand at $s = M_{\tau}^{2}$ suppresses the 
contribution from the dangerous segment close to the 
branch cut, where OPE has 
little chance to represent appropriately the function. 
(As a further means of suppressing  
bad influence of 
this segment, integral averaging smeared over an energy 
of the order $\Lambda_{{\rm QCD}}$ can be used.) 
Application of the estimates discussed in subsection 3.2 will 
give quantitative background to   
this qualitative assumption, specifying how reliable 
the expansion is at the points $|s|=M_{\tau}^2$ that 
are far both from the euclidean interval and from the 
immediate neighbourhood of the point $s=M_{\tau}^2$, 
thereby possibly modifying the accuracy of some of the estimates.  

The estimate of $\alpha_{s}$ and of the QCD condensates
from e$^{+}$e$^{-} \rightarrow I=1$  data was re-examined 
in \cite{Narison} using $\tau$-like inclusive process and 
QCD sum rules. The resulting values confirm previous sum rules
estimate based on stability criteria. Implication on the value
of $\alpha_{s}$ from $\tau$-decays gives 0.33 $\pm$ 0.03 as the
value of $\alpha_{s}(M_{\tau})$.  

Applications of the operator-product expansion to heavy 
quark physics (such as B and $\Lambda_{{\rm b}}$ decays, the 
problem of heavy-quark pole mass) are based on an extension 
of the idea of the operator-product expansion to a new 
situation, when the expansion is made in terms of operators 
in the heavy-quark effective field theory, in inverse powers of 
a quantity of the scale of the heavy quark mass $m_{{\rm Q}}$ 
\cite{Grinstein}. Heavy-quark field theory is, to leading order, 
a theory for light quarks in the field of a static colour source. 
Evidently, $m_{{\rm Q}}$, this key quantity, is not uniquely 
defined because quarks are unobservable; the so-called pole mass 
$m_{{\rm Q}}^{{\rm pole}}$ (i.e., the position of the pole in 
the quark propagator) is well-defined in each finite-order  
perturbation theory \cite {BSUV} but, as the authors %of \cite{BSUV} 
show,  no precise definition of the pole mass exists once 
nonperturbative effects are included. They argue that any 
consistent definition of this quantity suffers from an intrinsic 
uncertainty of order [pole mass times 
$\Lambda_{{\rm QCD}}/m_{\rm Q}$]. This circumstance can be described
in terms of an infrared renormalon of the type (\ref{IR}), which 
produces a factorial divergence of the perturbation series.  

As in the $\tau$ decay case, also in the case of the 
heavy-quark systems the approach starts from the
operator-product expansion in euclidean region, this time 
in the kinematic invariant $v.q$, where $v$ is the velocity 
of the heavy quark. Again, the problem leads to a contour 
integral of the type (\ref{Rtau}), which has a segment close 
to the cut, where the expansion is not safe, and it is 
expected that smearing over an energy of the order 
$\Lambda_{{\rm QCD}}$ would make valid the OPE computation 
of the differential distributions. 
%, including nonperturbative corrections. 
Results for semileptonic B and $\Lambda_{{\rm b}}$ decays 
are obtained in \cite{Grinstein}. 

The relation of the heavy quark mass to the expansion 
parameter is investigated in \cite{BB1}. It is shown that 
the heavy-quark effective theory suffers from a series of 
ultraviolet renormalons which are not Borel-summable. The 
heavy-quark effective theory should  be 
independent of the heavy-quark mass; since the $S$-matrix 
elements in a confining theory have no poles corresponding 
to a physical quark there is no natural choice of the 
expansion parameter.   
The ambiguity of perturbation series is related to an infrared 
renormalon in the pole mass and expresses the necessity of 
including an additional (and ambiguous) mass term.

The structure of renormalons in the heavy-quark effective theory
is examined in \cite{MarNeSa}. The authors show that the way in
which renormalons appear in effective theories is not universal,
but depends on the regularization scheme used to define the
effective theory. In the case of kinetic energy operator, they
find that the leading ultraviolet renormalon is absent in all
but one regularization schemes.
 
The authors of \cite{MarSach} develop a nonperturbative method 
for defining the higher-di\-men\-sional operators of the heavy quark 
effective theory such that the matrix elements are free of 
ambiguities due to the UV renormalon singularities. They define a 
"subtracted pole mass" %$m^{{\rm S}}_{{\rm Q}}$, 
from which the renormalon singularities 
are subtracted in a nonperturbative way and whose inverse can be
used as the expansion parameter. 
  
For a critical review of recent achievements in the heavy quark
theory see ref. \cite{BSU}.

%%%%%%%%%%%%%%%%%%%%%%%%

\section{Renormalization ambiguities}

Results of any finite-order calculation depend on the 
renormalization prescription, although the theory as a
whole must be invariant with respect to the 
renormalization group.  
This issue has been lively discussed in QCD 
since many 
years in connection with theoretical predictions of 
observable quantities. As the renormalization-group 
dependence of a finite-order result can never be fully 
removed, the effort has been focused on looking for such 
a renormalization prescription (RP) and such a method of 
approximation which would 
possibly minimize the dependence on the choice of the RP. 
The extent of ambiguities connected with 
renormalization-group invariance is often not fully 
appreciated; we shall therefore 
explain the main concepts and the actual origin of 
the ambiguities. In expounding the subject we shall 
follow the way outlined in ref. \cite{Chyla}; for
practical aspects of the renormalization scheme 
dependence of the perturbative QCD approximations 
see also \cite{ChylKa}, \cite{Pra3}. 

Let $r(Q)$ be a generic observable quantity that depends, 
for simplicity, on a single external momentum $Q$,  
\begin{equation}
r(Q) = a\left[ 1 + r_{1}a + r_{2} a^2 + . . . \right] .
\label{r}
\end{equation}
On the right-hand side, the $Q$-dependence of $r(Q)$ is 
hidden in a $Q$ dependence of the expansion coefficients
$r_{k}$. Besides this, both $a$ (the renormalized coupling 
constant, 
often called coupling parameter or couplant in this context, 
to emphasize its running character) and the coefficients 
$r_{k}$ depend on the renormalization prescription (RP), 
which consists of the choice of the renormalization 
scale $\mu$ (of dimension energy) and of the so-called 
renormalization scheme (RS). Thus,
\begin{eqnarray}
a = a(\mu, {\rm RS})  \nonumber \\
r_{k} = r_{k}(Q/\mu , {\rm RS}) .
\end{eqnarray}
The quantity $r(Q)$, on the other hand, should not (by
definition) depend on $\mu$ or RS, because it is an
observable quantity.\footnote{Strictly speaking, this 
statement is true in the fictitious world of 
perturbative physics. Genuinely independent is the sum 
$r(Q) + r_{_{NPt}}(Q)$, where the second term is the 
nonperturbative part. But such a statement has no predictive
power unless something is known about $r_{_{NPt}}(Q)$. The
standard approach is to assume that $r(Q)$ and 
$r_{_{NPt}}(Q)$ are separately independent of the
renormalization prescription.} 

One can distinguish the following stages of fixing the 
renormalization-group ambiguities.   

1. {\it Renormalization scale}. The dependence of the couplant $a$ 
on the scale $\mu$ is governed by the differential equation
\begin{equation}
\frac{{\rm d}a(\mu, {\rm RS})}{{\rm d}\ln \mu} \equiv \beta (a) 
= - ba^2 (\mu, {\rm RS})\left( 1+ca(\mu, {\rm RS}) + 
c_{2}a(\mu, {\rm RS}) + ...\right) .
\label{RGE}
\end{equation}
The choice of the renormalization scale consists in selecting,
when equation (\ref{RGE}) has been solved, the value of the
parameter $\mu$, with respect to which the theory is
invariant. 

2. {\it Renormalization convention}. While $b$ and $c$, the 
first two coefficients on the right hand side of (\ref{RGE}), 
are uniquely determined as functions of $N_{f}$  and $N_{c}$,  
\begin{eqnarray} 
b=(11N_{c}-2N_{f})/6  \nonumber \\  
c=(51N_{c}-19N_{f})/(11N_{c}-4N_{f}) \nonumber
\end{eqnarray}
(where $N_{f}$ and $N_{c}$ is the number of massless
quark species and the number of colours respectively), 
all the other coefficients $c_{i}$ are completely 
arbitrary and define the {\it renormalization 
convention}, see Table 3.  
One can consider different renormalization conventions 
(i.e., different choices of the $c_{i}$), among which 
't Hooft's convention \cite{'t Hooft}, in which all 
the $c_{i}$ vanish, is remarkable for its simplicity and 
is suitable for general considerations; in this convention, 
the right-hand side of (\ref{RGE}) consists of two terms,
while (\ref{r}) is generally an infinite series. One can
think of another special choice, in which (\ref{r}) is a
truncated expression and (\ref{RGE}) is an infinite series
(Grunberg's convention, effective charges approach,
\cite{GEF}).

\begin{table}
\begin{tabular}{llll}
{\bf Symbol} & {\bf Name} & {\bf to be
fixed} & {\bf what happens} \\[10pt]
R$_{{\rm scale}}$ & renormalization scale & $\mu$  
   & scale chosen \\
RC & renormalization convention  & $c_{i}$  & 
form of curves chosen \\ 
RS & renormalization scheme & $c_{i}$, 
$\Lambda_{{\rm RS}}$ & curve chosen\\
RP & renormalization prescription & $c_{i}$, 
$\Lambda_{{\rm RS}}$, $\mu$ & complete choice
\end{tabular}
\caption[]{{\bf Stages of renormalization fixing}:
renormalization is fully fixed by specifying the 
renormalization prescription}
\end{table}

3. {\it Choice of the renormalization scheme}.  
Once the  coefficients $c_{i}$ are chosen and some boundary 
condition on $a(\mu, {\rm RS})$ is specified, the equation 
(\ref{RGE}) can be solved. Concerning the 
different ways of specifying the boundary conditions on 
$a(\mu, {\rm RS})$, a popular way consists in selecting 
the scale parameter $\Lambda_{RS}$, 
defined by the following implicit equation\cite{PMS} for
the solution of the differential equation (\ref{RGE})
\begin{equation}
b \ln \frac{\mu}{\Lambda_{RS}} = \frac{1}{a} + c \ln
\frac{ca}{1+ca} + \int_{0}^{a} {\rm d}x \left[-\frac{1}{x^2
B^{(n)}(x)}+\frac{1}{x^{2}(1+cx)}\right] ,
\label{imp}
\end{equation} 
where $B^{(n)}(x) = 1+cx+c_{2}x^{2}+ . . . c_{n-1}x^{n-1}$.
The integrand on the right-hand side of (\ref{imp}) is 
regular at the lower integration limit and vanishes in 't 
Hooft's convention, in which all the $c_{i}$ vanish. 

According to (\ref{imp}), the choice of $\Lambda_{{\rm RS}}$ 
amounts to selecting a special curve among the solutions 
to (\ref{RGE}), which in 't Hooft's convention has a
particularly simple interpretation: $\Lambda_{RS}$ is that
value of $\mu$ for which $a(\mu, {\rm RS})$ becomes infinite. 
Equations similar to (\ref{imp}) govern also the dependence 
of the couplant $a(\mu, c_{i})$ on the parameters $c_{i}$; 
the equations contain the same fundamental $\beta$ function 
and introduce no additional ambiguity.

To sum up, the full size of ambiguities connected with 
renormalization group invariance involves the choice of the 
renormalization scale $\mu$, the choice of the values of 
the $c_{i}$-coefficients, and the choice of the boundary 
condition on $a(\mu, {\rm RS})$ by specifying the 
renormalization scheme.  
A survey of degrees of renormalization
fixing is given in Table 3; let me just point out 
that the terminology (convention, scheme, prescription...), 
although widely adopted at present, is  sometimes used in a
different way.  

The explicit dependence of the expansion coefficients
$r_{k}$ on the renormalization parameters $\mu$,
$\Lambda_{{\rm RS}}$ and the $c_i$ is given by the
condition that $r^{(N)}(Q/\mu, {\rm RS})$, the sum of the
first $N$ terms of the right-hand side of (\ref{r}),
\begin{equation}
r^{(N)}(Q/\mu, {\rm RS}) = \sum_{k=0}^{N-1} r_{k}(Q/\mu, {\rm RS})
a^{k+1}(\mu, {\rm RS}) , \nonumber
\end{equation} 
when differentiated once with respect to $\ln \mu$, $c_{i}$ 
and $\ln \Lambda_{{\rm RS}}$ respectively, should be of 
the order $O(a^{N+1})$ (note that, in the latter case, it is 
not the value of $\Lambda_{{\rm RS}}$ what varies at fixed 
RS, but the change consists in a transition from one 
renormalization scheme to another).
This allows us to introduce the renormalization-group
invariants $\rho_{i}$, $i=1,2,3,...$ and represent the 
coefficients $r_{k}$ in terms of them in the following 
form 
\begin{eqnarray} 
r_{1}(Q/\mu, {\rm RS}) = b \ln\frac{\mu}{\Lambda_{{\rm RS}}}
- \rho_{1}(Q/\Lambda_{{\rm RS}})  \nonumber \\
r_{2}(Q/\mu, {\rm RS}) = \rho_{2}-c_{2}+r_{1}^{2}+c
r_{1} ,
\label{chyla}
\end{eqnarray}
etc., where only $\rho_{1}$ depends on $Q$, while the
$\rho_{i}$, $i=2,3,...$ are numbers, the dependence of
$\rho_{1}$ on $Q$ being obtained from (\ref{chyla}) by
setting $\mu = Q$.\footnote{It might seem paradoxical that 
the function $\rho_{1}$, while being a renormalization-group
invariant, is represented here as a function of
$Q/\Lambda_{RS}$, where  $\Lambda_{RS}$ is no invariant. 
As $\rho_{1}$ is dimensionless, its $Q$-dependence %occurs 
takes place only through dimensionless quantities. 
Then, to preserve invariance, the symbol $\rho_{1}$ 
must represent  different function prescriptions 
for different schemes.} 

Thus, although the expansion coefficients $r_i$ and $c_i$
are RP-dependent, certain combinations of them, the
function $\rho_{1}$ and the numbers $\rho_i$,
$i=2,3,...$, are RP-invariants. Their knowledge is 
relevant for the the extraction of QCD observables from
experiment (see, e.g., \cite{Chyla, ChylKa, BM2, PRa1, PRa2}).
There are two different sources of uncertainties in the
theoretical determination of a QCD quantity, one being 
connected with higher-order (uncalculated) perturbative 
corrections, the other with the renormalization
scheme and scale ambiguity. Since the exact expression
(perturbative sum) must be RG invariant (see the
footnote 6), there is a widespread belief that
a weak RG dependence is a signal of a small large-order
correction, i.e., that the strength of the RG dependence
of a particular observable is related to the size of its
higher-order corrections which, by this, can be minimized
by choosing the renormalization prescription that has the 
weakest dependence on the RG parameters.

%%%%%%%%%%%%%%%%%%%%%%%%%%%%%%%%%
\section{Concluding remarks}

A typical feature of the present status of the QCD perturbative
corrections is the trend to avoid explicit calculation of 
higher-order corrections and, instead, to improve the result 
by making use of some additional information. 
Such information may be, for instance, the 
renormalization group invariance (which allows us to 
introduce the running coupling constant instead of the fixed 
one), analyticity, or the structure of singularities in the 
Borel plane. 

As was already mentioned, the notion of renormalon, originally 
introduced and used to investigate interesting mathematical 
models, is now widely believed to have concrete background 
in physical phenomena. This wide-spread opinion meets also with 
scepticism, which argues that some singularities in the Borel 
plane might be an attribute of the renormalization scheme used 
\cite{Krasnikov}. 

It seems that the present effort in further developing the 
idea of renormalon and the corresponding formalism will be 
helpful in finding a language %//framework 
appropriate for supplying the non-perturbative sector with 
new physical ideas.  
Generalizations of the scale-setting procedures 
become valuable by implementing new physical information 
without the need to calculate higher-order perturbative 
corrections 
(which is not only cumbersome but also doubtful, if the series 
is divergent or convergent with a bad rate), by using some 
additional, perturbatively independent, information. This idea 
is not new, appearing in theoretical physics whenever technical 
difficulties force one to look for methods allowing the 
exploitation of all information on the system, including 
whatever is not adequately taken into account by the existing 
formalism. 

This additional information is needed not only   
whenever one is unable to calculate higher-order
corrections. It seems evident that perturbative quantum 
chromodynamics is not self-sufficient on principle, even if 
we had calculated all the perturbative coefficients of any 
order. The asymptotic perturbative series calls for 
new dynamical input. The operator product
expansion, however bringing valuable physical contribution,
is not sufficient either, being as a power series connected 
with a similar ambiguity. It is intriguing to have the 
missing input supplied by the general principles of the 
$S$-matrix theory, among which analyticity occupies the 
central position. This is not only due 
to its rich physical content, but 
also because, as was pointed out in subsection 1.2, the
$S$-matrix possesses the standard analyticity properties
even for gauge theories with confinement \cite{Oeh}. 
This fact distinguishes all summation methods that are able to 
make use of analyticity, as additional input, to remove the
ambiguity of a power expansion in defining the expanded
function. 

The optimal summation method may be different from
whatever we now know. It may be Borel, some of its
generalizations, Pad\'e, but also something very different. The
criterion should not be mathematical simplicity or elegance of
the method or result, but %economy and 
adequateness with which the method in question is able to 
glean all the physical information that the standard 
conventional field (and perturbation) theory failed to take 
into account and process.   
 
Summarizing, one can say that the difficulties of
the present quantum chromodynamics are, first, of technical
nature (technical impossibility of calculating higher-order
perturbative coefficients) and, second, of principal nature
(ambiguities of asymptotic power expansions, renormalization
ambiguities). All of them point to insufficiency of the
current state of QCD and have in common the feature that they
represent a challenge to nonperturbative quantum chromodynamics, 
without which no essential progress seems to be possible.
 
\vskip 0.6cm

\noindent{\bf Acknowledgements}

This paper is dedicated to the memory of 
Ryszard R\c{a}czka, my excellent teacher, colleague and 
friend. I am indebted to Chris Maxwell, Stephan Narison and Ivo 
Vrko\v{c} for stimulating discussions, and to Martin Beneke and 
Ji\v{r}\'{i} Ch\'{y}la for carefully reading the manuscript and
valuable comments. This work has 
been supported in part by GACR and GAAV (Czech Republic) under 
grant Nos. 202/96/1616 and A1010603 respectively.

\end{document}